\documentclass[12pt,letter]{article}
\usepackage[left=1in,right=1in,top=1in,bottom=1in]{geometry}
\usepackage{comment}
\usepackage[ruled,vlined]{algorithm2e}
\usepackage{amsfonts}
\usepackage{amsmath}
\usepackage{amssymb}
\usepackage{amsthm}
\usepackage{array}
\usepackage{authblk}
\usepackage{bm}
\usepackage{booktabs}
\usepackage{caption}
\usepackage{color}
\usepackage{graphicx}
\usepackage[colorlinks,linkcolor=blue,citecolor=blue,urlcolor=blue]{hyperref} 
\usepackage{multirow}
\usepackage[authoryear,semicolon]{natbib}
\usepackage{physics}
\usepackage{siunitx}
\usepackage{standalone}
\usepackage{setspace}
\usepackage{subfigure}
\usepackage{tikz}
\usepackage[normalem]{ulem}
\usepackage{url}
\usepackage{lineno}
\usepackage{rotating}
\usepackage{gensymb}
\allowdisplaybreaks
\usepackage{placeins} 

\newcommand*\patchAmsMathEnvironmentForLineno[1]{%
  \expandafter\let\csname old#1\expandafter\endcsname\csname #1\endcsname
  \expandafter\let\csname oldend#1\expandafter\endcsname\csname end#1\endcsname
  \renewenvironment{#1}%
     {\linenomath\csname old#1\endcsname}%
     {\csname oldend#1\endcsname\endlinenomath}}%
\newcommand*\patchBothAmsMathEnvironmentsForLineno[1]{%
  \patchAmsMathEnvironmentForLineno{#1}%
  \patchAmsMathEnvironmentForLineno{#1*}}%

\AtBeginDocument{%
\patchBothAmsMathEnvironmentsForLineno{equation}%
\patchBothAmsMathEnvironmentsForLineno{align}
\patchBothAmsMathEnvironmentsForLineno{flalign}%
\patchBothAmsMathEnvironmentsForLineno{alignat}%
\patchBothAmsMathEnvironmentsForLineno{gather}%
\patchBothAmsMathEnvironmentsForLineno{multline}%
}

\newcolumntype{x}[1]{>{\centering\arraybackslash\hspace{0pt}}p{#1}}
\newcommand{\Var}{\text{Var}}

\newcommand{\txd}{\text{d}}

\newcommand{\N}{\mathcal{N}}

\newcommand{\alphavec}{\boldsymbol{\alpha}}
\newcommand{\betavec}{\boldsymbol{\beta}}

\newcommand{\phivec}{\boldsymbol{\phi}}
\newcommand{\gammavec}{\boldsymbol{\gamma}}

\newcommand{\thetavec}{\boldsymbol{\theta}}
\newcommand{\xivec}{\boldsymbol{\xi}}
\newcommand{\zetavec}{\boldsymbol{\zeta}}
\newcommand{\zerovec}{\boldsymbol{0}}

\newcommand{\Sigmavec}{\boldsymbol{\Sigma}}

\newcommand{\bvec}{\mathbf{b}}
\newcommand{\cvec}{\mathbf{c}}

\newcommand{\svec}{\mathbf{s}}

\newcommand{\xvec}{\mathbf{x}}

\newcommand{\Ivec}{\mathbf{I}}

\theoremstyle{definition}

\begin{document}

\title{\textbf{\Large{A Single Index Approach to Integrated Species Distribution Modeling for Fisheries Abundance Data}}}  
  \author[1]{\normalsize{Quan Vu}}
  \author[1]{\normalsize{Francis K. C. Hui}}
    \author[1]{\normalsize{A. H. Welsh}}
  \author[2]{\normalsize{Samuel Muller}}
  \author[3]{\normalsize{Eva Cantoni}}
  \author[4]{\normalsize{Christopher R. Haak}}
  \affil[1]{\small{Research School of Finance, Actuarial Studies and Statistics, The Australian National University, Australia}}
  \affil[2]{\small{Faculty of Science and Engineering, Macquarie University, Australia}}
  \affil[3]{\small{Geneva School of Economics and Management, University of Geneva, Switzerland}}
  \affil[4]{\small{Urban Coast Institute, Monmouth University, United States}}
  \date{}
  \maketitle

\begin{abstract}

In fisheries ecology, species abundance data are often collected by multiple surveys, each with unique characteristics. This article is motivated by a dataset of Atlantic sea scallop abundance records along the northeast coast of the United States, collected from two bottom trawl surveys which cover a larger spatial domain but have low catch efficiency, and a dredge survey which is more efficient but more bounded in domain.
Over the past decade, integrated species distribution models (ISDMs) that include common environmental effects along with correlated survey-specific spatial fields have been used to incorporate information from multiple surveys. While flexible, ISDMs can be susceptible to overfitting,
which can complicate interpretability of the shared environmental effects, and potentially lead to poor predictive performance. 
To overcome these drawbacks, we introduce a novel single index ISDM, built from a single index (with spatial random effects) that represents a latent measure of the true species distribution, and survey-specific catch efficiency functions which map the single index to the survey-specific expected catch.
In this article, these functions are constructed via logistic functions or semiparametric spline-based functions.
Simulations and application to the motivating sea scallop abundance data demonstrate that the proposed single index ISDM offers more meaningful interpretations of the environmental effects and survey catch efficiency differences, while achieving similar to or better predictive performance than existing ISDMs. 

\end{abstract}

\section{Introduction} \label{sec:intro}

In ecology, it is becoming increasingly common to access and analyze datasets from multiple sources that can be quite different in structure and properties \citep{isaac2020data, rufener2021bridging}. 
For example, this article is motivated by fisheries ecology, specifically, counts of Atlantic sea scallop (\emph{Placopecten magellanicus}) abundance along the United States northeast shelf, collected from three different National Oceanic and Atmospheric Administration (NOAA) surveys that vary in their spatial footprints, sampling protocols, and fishing gear types. These surveys produce count records that differ in their relationships with (or representation of) the `true' underlying species densities. 

A common goal in analyzing species abundance data from multiple sources (note in this article we use the terms `surveys' and `sources' interchangeably), and indeed the goal of this article, is to develop a statistical model that yields interpretable insights into environmental and habitat-related factors. Furthermore, by integrating data across multiple sources, we can expect to achieve enhanced predictive performance through the sharing of information. 
In contrast, the simpler approach of fitting independent models to each source fails to account for the fact that species records from various sources fundamentally arise from the same underlying abundance distribution. 

As a means of systemically combining data from multiple surveys, integrated species distribution models \citep[ISDMs,][]{fletcher2019practical, gruss2023spatially, simmonds2020more}
have emerged over the past decade, based on the fundamental idea that the distributions of observations from different data sources are assumed to share certain components, most commonly in the linear predictors. 
For example, the linear predictors from different sources can share the same fixed and/or random effects, while other effects  unique to each source are included to account for survey-specific characteristics \citep[e.g.,][]{rufener2021bridging}. 
One popular version of ISDMs, particularly for spatially indexed abundance data, involves fitting a generalized linear mixed model (GLMM) where a spatial field is included for one of the data sources, often viewed in fisheries ecology as the ``reference survey''. The spatial field for every other data source is then formed from the sum of this first spatial field with another field specific to that source; such structure allows for shared spatial information while also accommodating source-specific variation \citep[see][for details of such a formulation]{dovers2024fast, gruss2023spatially, simmonds2020more}. Finally, fixed effects representing environmental and habitat factors are typically shared across all sources. In this article, we refer to such models as additive field ISDMs.


From an interpretation standpoint, the additive field ISDM attributes differences in catch efficiency between data sources solely to differences in spatial random effects. 
Such models therefore do not explicitly consider the ecological and sampling processes responsible for the differences between the multiple surveys. 
For instance, in the case of the motivating scallop abundance data, there is evidence dredge capture efficiency varies across different substrate types but also with the density of the scallops in the tow path \citep[see][for more ecological details behind this]{delargy2023global}.
Statistically speaking, we thus expect that as the true species abundance increases the recorded abundance for all surveys should also increase, but the rate and ``shape'' of this increase will depend on the catch efficiency of each survey. 
The additive field ISDM does not explicitly reflect such catch efficiency, instead relying on correlated spatial fields as an omnibus means of accounting for differences between surveys. One consequence is that ISDMs can end up being too flexible, and as we will show later in Section \ref{sec:application_results} this can lead to adverse performance in the scallop abundance data.

In this article, we propose a new single index integrated species distribution model (siISDM) 
which aims to address the drawbacks mentioned above by constructing a model that more explicitly accounts for the heterogeneous catch efficiency across surveys, thus facilitating better interpretation and prediction. As the name suggests, the model works by constructing a single index, comprising fixed and random effects capturing environmental and habitat factors and residual spatial correlations, which represents a measure of the true species distribution. Importantly, this single index is shared across all surveys. Differences in catch efficiency between surveys are then captured through so-called survey-specific catch efficiency functions (CEFs), which map the single index to the mean abundance from each survey. 
Ecologically, by explicitly constructing CEFs, the siISDM offers a novel understanding of the efficiency of different sampling gears, and how it varies as a result of underlying density.
We constrain these CEFs to be monotonically increasing, to reflect the idea that if true abundance increases, the mean abundance in each survey should in general also increase. Otherwise, these functions are estimated in a relatively flexible manner e.g., via a parametric four-parameter logistic function \citep{ratkowsky1986choosing}, or a semiparametric approach based on monotonic splines \citep{ramsay1988monotone}. We estimate siISDMs using maximum likelihood estimation via Template Model Builder \citep[TMB,][]{osgood2023statistical}, which offers computationally scalable fitting courtesy of automatic differentiation techniques coupled with a Laplace approximation to approximate the marginal log-likelihood function.

Simulations and an application to the motivating multi-survey scallop abundance data demonstrate that siISDMs can provide meaningful estimates of how environmental factors affect the true species distribution, while the estimated survey-specific CEFs offer insights into the relative catch efficiency of the surveys. Furthermore, siISDMs can offer similar or better predictive performance across surveys when compared with both independent SDMs as well as some current ISDM approaches including the additive field ISDM. Particularly with the scallop abundance case study, where species' recorded prevalence can be low, the strong predictive performance of siISDMs suggests it strikes a strong balance between ecological interpretability and flexibility.

The remainder of this article is organized as follows. In Section \ref{sec:data}, we offer some more details for the three surveys forming the motivating scallop abundance data. Section \ref{sec:isdm} reviews independent SDMs and ISDMs, with a particular focus on the additive field formulation. Section \ref{sec:single_index_model} introduces the proposed siISDM along with details on estimation and prediction. Section \ref{sec:simulation} presents simulation studies to illustrate the performance of siISDMs, while Section \ref{sec:application_results} analyzes the motivating scallop abundance data. Finally, Section \ref{sec:discussion} gives some concluding remarks.

\section{Scallop abundance data} \label{sec:data}
Atlantic sea scallops (\emph{Placopecten magellanicus}) are a type of mollusca found naturally across the United States Northeast Shelf from Newfoundland to North Carolina. 
This article focuses on the distribution of Atlantic sea scallops between 35.0$\degree$N and 45.0$\degree$N latitude, and between 77.5$\degree$W and 65.0$\degree$W longitude. The data is sourced from three distinct surveys run by National Oceanic and Atmospheric Administration (NOAA): 
\begin{itemize}
    \item \underline{Bottom trawl surveys:} The Northeast Fisheries Science Center (NEFSC) bottom trawl survey 
    \citep{politis2014northeast}  
    is a broad-scale, non-selective multi-species trawl survey that provides information on relative abundances and distributions for a wide variety of marine organisms inhabiting the Northeast Shelf Large Marine Ecosystem. The trawl survey employs a mesh trawl net
whose footrope or sweep (i.e., bottom of the net) is equipped with rollers that suspend it just above the seabed. This reduces capture probability for benthic invertebrates resting directly on the substrate such as scallops. 
Importantly, in 2009 the bottom trawl survey transitioned to a more modern vessel and gear with improved bottom contact, resulting in generally greater catch efficiency for bottom-associated taxa \citep{miller2010estimation}. With this in mind, we consider the two time periods, namely pre-2009 and 2009-onward, as two distinct trawl surveys. There are 5,933 observations in the trawl survey from 2000 to 2008, and 8,877 observations in the trawl survey from 2009 to 2022.

\item \underline{Scallop dredge survey:}
The NEFSC scallop dredge survey
\citep{hart2006long}
is a targeted single species survey with the principal goal of informing stock assessments for Atlantic sea scallops. 
It focuses on regions where commercial fishing tends to be more concentrated, including parts of Georges Bank and the Mid-Atlantic Bight, and is contained entirely within the footprint of the trawl survey. 
Designed to target scallops, the leading edge of the metal dredge digs into the seabed, resulting in higher catch efficiency and thus higher observed catch rates.
There are 6,468 observations from the dredge survey from 2000 to 2022.
\end{itemize}

\begin{figure}[t]
\centering
\includegraphics[width = \textwidth]{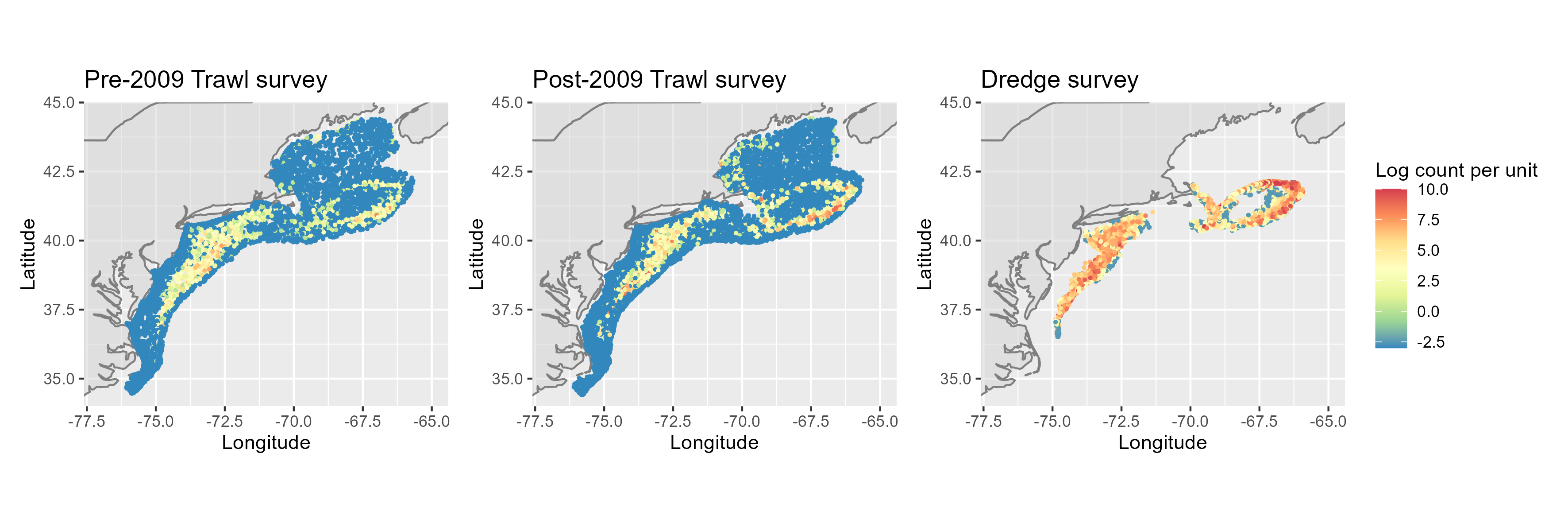}
\caption{Log of count of Atlantic sea scallops per area unit from the trawl and dredge surveys. Zero count is represented by dark blue in the figures.
}
\label{fig:obs}
\end{figure}

The trawl and dredge surveys are scientific, fishery-independent surveys employing stratified random sampling designs with strata based on water depth and latitude or geographic region \citep{hart2006long}. 
While the dredge survey operates within a spatial subdomain of the broader area covered by the trawls, it does so systematically and without substantial bias in the mean conditions sampled.
Figure \ref{fig:obs} presents spatial plots of all three surveys, from which the aforementioned differences in the spatial sampling region of the surveys are evident.
Moreover, as hinted at above, comparing across the three plots reveals that the dredge survey tends to provide higher recorded counts than the two bottom trawl surveys.
Put another way, although the three surveys are attempting to capture the same underlying true abundance distribution, differences in gear and sampling protocol 
mean that each data source has its own mapping from the true species distribution to the realized recorded abundances. 

Turning to environmental and habitat factors, because scallops are a mainly sessile (immobile) benthic species, we selected environmental covariates that reflect long-term conditions at or near the seabed. Specifically, we consider the following nine measures of environmental and local habitat conditions: annual mean bottom salinity, annual mean bottom temperature, 95th quantile of bottom (hydrodynamic) stress, bottom depth or bathymetry, tidal current velocity, chlorophyll-a concentration, topographic complexity, bathymetric position index (BPI), and sediment grain size. All variables were continuous and standardized to have mean zero and variance one prior to analysis. We provide more details on how each of these covariates were recorded, including a visualization of them in Supplementary Material S1. 
Finally, for all three surveys, tow distances and dredge width or trawl wing-spread were used to compute the swept area for each individual tow \citep[see][for further details]{nefsc201050th,politis2014northeast}; these will be included as offset terms as part of the analysis below.

\section{Integrated species distribution models} \label{sec:isdm}
In this section, we first review independent species distribution models i.e., fitting a separate model to each survey, before formulating the additive constant ISDM as well as the more general additive field ISDM \citep[e.g.,][]{dovers2024fast, gruss2023spatially, simmonds2020more}. The last of these models in particular has grown in popularity over the past decade for analyzing multiple-survey species abundance data.

All the models introduced in this section, as well as the siISDM proposed in the next section, start with the same setup for the distribution of the response. Specifically, consider a set of $m$ distinct data sources (note $m = 3$ surveys in the context of the scallop abundance data), where for source $j = 1, \dots, m$ we record a vector of $n_j$ observations of species abundance $\{ y_j(\svec_1), \dots, y_j(\svec_{n_j})  \}^\top$ at spatial locations in some domain $\svec_i \in \mathcal{D}_j; i = 1,\ldots,n_j$. Additionally, we observe a set of $p$ covariates $\xvec(\svec_i) = \{x_{i1}(\svec_i), \dots, x_{ip}(\svec_i)\}^{\top}$ at each location, where $x_{i1} = 1$ to represent an intercept term. Note all $p$ covariates usually are available for each observed data point in (fisheries) ecology.

Next, for survey $j$ we assume the $y_j(\svec_i)$'s follow a distribution characterized by the mean parameter $\mu_j(\svec_i)$, along with one or more survey-specific dispersion parameters $\bm{\phi}_j$ which may be known or also require estimation. That is,  $y_j(\svec_i) \sim f\{y_{j}(\svec_i) \mid \mu_j(\svec_i), \bm{\phi}_j\}$. Note that in general the distribution $f(\cdot \mid \cdot, \cdot)$ can also be survey-specific; this is especially relevant if the different data sources produce different response types \citep[e.g.,][]{dovers2024fast}. However, given the three surveys comprising the motivating scallop abundance records all produce count responses, then we will focus on the case where the same distributional form is assumed for all surveys.  
In particular, given the overdispersed nature of the count data as suggested in Figure \ref{fig:obs}, then we assume $f(\cdot \mid \cdot, \cdot)$ follows a negative binomial distribution with quadratic mean-variance relationship $\Var\{y_j(\svec_i)\} = \mu_j(\svec_i) + \phi_j \mu^2_j(\svec_i)$ \citep[e.g.,][]{stoklosa2022overview}, where
the log link is used to relate the mean to the survey-specific linear predictor,
\begin{equation} \label{eqn:basicmeanmodel}
\log\{ \mu_j(\svec_i) \} = \eta_j(\svec_i), ~ j = 1, \dots, m; \; i = 1,\ldots,n_j.
\end{equation}


\subsection{Existing models} \label{subsec:existing_models}
A starting point to analyze multi-source data is the independent species distribution model or independent SDM. As the name suggests, this fits a separate GLMM to each data source, 
\begin{align*}
\text{\underline{Independent SDM:}} \quad \eta_j(\svec_i) = \xvec(\svec_i)^{\top} \betavec_j + u_j(\svec_i), ~ j = 1, \dots, m;  \; i = 1,\ldots,n_j,
\end{align*}
where the covariate effects $\betavec_j = (\beta_{j1},\ldots,\beta_{jp})^\top$ are survey-specific and the spatial random effect $u_j(\cdot)$ are assumed to be independent across the $m$ surveys; we provide more details regarding how the spatial random effect is set up towards the end of Section \ref{sec:single_index_model}. Offset terms can also be added to the $\eta_j(\cdot)$'s to account for differences in sampling effort, e.g., area swept, although for simplicity we omit the notation for this throughout the article.

The independent SDM is parametrized by entirely separate sets of fixed and random effects for each data source. This prevents it from being readily interpretable, as it is not possible to infer how environmental and habitat factor affect the underlying true species distribution, nor to infer the difference in catch efficiencies between the surveys. Moreover, because it does not borrow strength across the data sources, it is expected to have poorer predictive performance than models which formally integrate the data sources \citep[e.g.,][]{dovers2024fast}; see also the results in Sections \ref{sec:simulation} and \ref{sec:application_results} respectively for explicit evidence of this.

The additive field ISDM overcomes some of the problems in the independent SDM approach by sharing/correlating components across different survey-specific linear predictors. Specifically, 
\begin{align*}
\text{\underline{Additive field ISDM:}} \quad \eta_j(\svec_i) &= \xvec(\svec_i)^{\top} \betavec + u_j(\svec_i), ~ j = 1, \dots, m;  \; i = 1,\ldots,n_j \\
u_j(\svec_i) &= u_1(\svec_i) + \tilde u_j(\svec_i), j = 2, \dots, m,
\end{align*}
from which we see that: 1) the covariate effects are assumed to be shared across the surveys i.e., $\betavec_1 = \dots = \betavec_m = \betavec$, where $\betavec$ is now the shared effects across all surveys; 2) there are correlations between spatial random effects across data sources. In particular, these correlations are induced by first selecting a survey to be the so-called ``reference survey'' and assigning it the spatial random effect $u_1(\cdot)$. The quantities $\tilde u_j(\cdot), j = 2, \dots, m$, which are all independent of each other and of $u_1(\cdot)$, then explain differences in the residual spatial correlations of the remaining surveys relative to the reference survey, analogous to how dummy variables work when setting up factors as covariates in a regression model.
Note a consequence of this additive field ISDM is that the total spatial random effect for the non-reference surveys will always have a larger variance than the spatial random effect for the reference survey. This means there needs to be a strong belief that the latter inherently has the smallest variance among all $m$ surveys. 

Finally, the additive constant ISDM shares the spatial random effects across surveys i.e., $u_j(\svec_i) = u(\svec_i), j = 1, \dots, m$, along with all the covariate effects \emph{except} the intercept. That is,
\begin{align*}
\text{\underline{Additive constant ISDM:}} \quad  \eta_j(\svec_i) = \beta_{1j} + \xvec(\svec_i)^{\top}_{-1} \betavec_{-1} + u(\svec_i), ~ j = 1, \dots, m;  \; i = 1,\ldots,n_j,
\end{align*}
where only the intercepts are survey-specific, and $\xvec(\svec_i)_{-1}$ and $\betavec_{-1}$ denote the vectors $\xvec(\svec_i)$ and $\betavec$ respectively with the corresponding first elements removed. The additive constant ISDM thus assumes differences in catch efficiency are constant in space. 
Moreover, by reparametrizing $\beta_{1j} = \beta_{11} + \tilde{\beta}_{1j}$ for $j = 2,\ldots,m$, we can view this model as a restricted case of the additive field ISDM.

\section{Single index integrated species distribution models} \label{sec:single_index_model}
As we explained in Sections \ref{sec:intro} and \ref{subsec:existing_models}, there are several drawbacks to existing ISDMs both statistically and interpretively. To address these drawbacks, we propose the single index ISDM (siISDM) which more explicitly attempts to quantify the relationship between the true species distribution that is common across all surveys, and the heterogeneous survey-specific recorded abundances that arise due to differences in catch efficiency.

As the name suggests, in an siISDM the log mean abundance $\eta_j(\svec_i)$ for the $j$th survey in \eqref{eqn:basicmeanmodel} is formulated to be a function of a single index $\kappa(\svec_i)$ which is common across the surveys. Specifically, 
\begin{subequations}
\begin{align} 
\eta_j(\svec_i) &= h_j \{ \kappa(\svec_i) \} , ~ j = 1, \dots, m;  \; i = 1,\ldots,n_j \label{eq:mean_indexmodel} \\
\kappa(\svec_i) &= \xvec(\svec_i)^{\top} \betavec + u(\svec_i). \label{eq:single_index_eta}
\end{align}
\end{subequations}
We refer to the functions $h_j(\cdot)$ in \eqref{eq:mean_indexmodel} as catch efficiency functions or CEFs, as they play the role of mapping the single index to the (log) mean abundance of each survey. We constrain the $h_j(\cdot)$'s to be monotonically increasing functions, such that a larger value of the single index corresponds to a larger expected count of the species. This is consistent with the idea that as the true species abundance increases, the recorded abundance in each data source should also increase. But otherwise, we allow the CEFs to vary in shape so as to reflect the distinct characteristics (e.g., gear and sampling protocol in the case of the motivating scallop abundance data) of each survey. We provide two approaches to estimating CEFs in Section \ref{subsection:catch_efficiency_functions}.

In \eqref{eq:single_index_eta}, we model the single index as a linear combination of the covariates $\xvec(\svec)$ plus a spatial random effect $u(\cdot)$, although see later for constraints to ensure parameter identifiability. Importantly, we interpret $\kappa(\svec)$ as an index of the underlying true species abundance distribution common to all data sources. That is, because $\kappa(\svec)$ is connected to a scale of mean abundance $\mu_j(\cdot)$ via a non-trivial convolution of $h_j(\cdot)$ and $\exp(\cdot)$, we view it as a relative measure of the true species distribution. This means the coefficients $\betavec$ and spatial random effect $u(\cdot)$ have to be interpreted in a relative sense e.g., in terms of their signs and relative magnitudes with respect to each other, but nevertheless they offer an explicit quantification of how environmental and habitat factors (along with residual spatial correlations) drive the true species distribution after accounting for heterogeneous catch efficiencies. Also, because $\kappa(\svec)$ is regarded as a measure of the true species density, then there is motivation to include non-linear functions of the elements of $\xvec(\svec)$, and indeed in our application to the motivating scallop abundance data in Section \ref{sec:application_results} we include both linear and quadratic terms for all nine measures of environment and local habitat to reflect the idea of an environmental niche \citep{austin2002spatial,van2021model}. 

Statistically, the siISDM shares similarities with both  (monotonic) single index models \citep[e.g.,][]{hardle1993optimal,foster2013variable,balabdaoui2019least}, and the single index model for multivariate responses \citep{feng2021sparse}. However, we stress that while this connection allows us to potentially leverage some of the existing literature for estimating and performing inference in the siISDM, e.g., estimating the functions $h_j(\cdot)$ using splines \citep{wang2009spline}, their motivations are very different. 
Traditional single index models are typically used for sufficient dimension reduction, noting that almost always the index only involve including covariates as linear terms.
By contrast, the siISDM is motivated by specific ecological interpretations regarding both the index of true species distribution and the CEFs.

Equations \eqref{eq:mean_indexmodel}--\eqref{eq:single_index_eta} require additional constraints to ensure the siISDM is identifiable. For instance, for a given constant $c \ne 0$, we could define $\check{\betavec} = c\betavec$, $\check{u}(\cdot) = c u(\cdot)$, and $\check{h_j}(\cdot) = c^{-1} h_j(\cdot)$, and the resulting linear predictors 
would be the same as those defined by $\betavec, u(\cdot)$, and $h_j(\cdot)$. To overcome this, we omit the intercept term in $\xvec(\svec)$ and set the first slope coefficient equal to one i.e., $\beta_{j1} = 1$. Such constraints are similar to those employed in single index models more generally \citep[e.g.,][]{foster2013variable,luo2016single}.

\subsection{Construction of the catch efficiency functions} \label{subsection:catch_efficiency_functions}
In this article, we examine two approaches to constructing the CEFs in \eqref{eq:mean_indexmodel}. The first is a parametric choice based on the four-parameter logistic function \citep{ratkowsky1986choosing},
\begin{equation}\label{eq:logistic_link_fn}
h_j\{\kappa(\svec)\} =  \frac{\gamma_{j1} }{1 + \exp{- \gamma_{j4} \kappa(\svec) + \gamma_{j3} } } + \gamma_{j2}; ~ j = 1,\ldots,m,
\end{equation}
where $\gammavec_j = ( \gamma_{j1}, \gamma_{j2}, \gamma_{j3}, \gamma_{j4} )^{\top}$ denote the vector of parameters characterizing the $j$th CEFs.
Equation \eqref{eq:logistic_link_fn} is a relatively simple class of monotonically increasing functions that allows straightforward computation during the model fitting process. 

As a second, more flexible approach to constructing the CEFs, we consider monotonic I-splines \citep{ramsay1988monotone} which can be embedded into the siISDM as 
\begin{align}\label{eq:spline_link_fn}
h_j\{\kappa(\svec)\} &= \zetavec[ G_{\bm{\tau}_j}\{\kappa(\svec)\} ]^{\top} \gammavec_{j},
\end{align}
where $G_{\bm{\tau}_j}\{x\}$ transforms the single index from the real line to the interval $(0,1)$ and is needed given the domain of the I-splines functions, $\zetavec(\cdot)$ denotes a vector of I-spline basis functions, 
and $\gammavec_{j}$ is the corresponding vector of basis function coefficients. We refer to \citet{ramsay1988monotone} as well as Supplementary Material S2.2 
to provide more details on the construction of the I-splines, acknowledging that other semiparametric approaches to constructing monotonic CEFs could be explored in future research \citep[see also the recent review of][]{dumbgen2024shape}.
Note because the range of the single index $\kappa(\svec)$ in \eqref{eq:single_index_eta} is not known in advance, then we allow $G_{\bm{\tau}_j}(x)$ in \eqref{eq:spline_link_fn} to further depend on a set of unknown source-specific location and scale parameters. For example, in this article we use the simple form $G_{\bm{\tau}_j}(x) = [1 + \exp{- \tau_{j1} \kappa(\svec) + \tau_{j0}}]^{-1}$, although other potentially more nonparametric transformations are possible
\citep[e.g., see][in the context of functional regression]{mclean2014functional}.  
The semiparametric approach to constructing CEFs permits a wider range of shapes compared with the four-parameter logistic function of \eqref{eq:logistic_link_fn}, but comes at the cost of greater computational challenges in estimating $(\gammavec_j^\top, \bm{\tau}_j^\top)^\top$.

\subsection{Construction of the spatial random effects} \label{subsection:spatial_random_effects}

We now turn to the construction of the spatial random effects $u(\cdot)$ in the siISDM in \eqref{eq:single_index_eta}, as well as in the SDMs reviewed in Section \ref{subsec:existing_models}. Specifically, a common approach to setting up spatial random effects is via some form of Gaussian process 
\citep{cressie2011statistics}. However, for large datasets such as the motivating scallop abundance dataset records where the total number of spatial locations across three surveys is $N = \sum_{j=1}^3 n_j = 21,278$, we 
require a more computationally efficient alternative. In this article, we employ fixed rank kriging \citep{cressie2008fixed, sainsbury2024modeling} which constructs the spatial random effect using basis functions \citep[see e.g.,][and references therein for some other potential alternatives]{heaton2019case}. Specifically, we assume $u(\svec_i) = \bvec(\svec_i)^\top \alphavec + \xi(\svec_i)$ where $\bvec(\svec_i) = \{b_1(\svec_i), \dots, b_B(\svec_i)\}^{\top}$ denotes a vector of $B \ll N$ bi-square basis functions evaluated at the spatial location $\svec_i$ i.e., $ b_k(\svec_i) = \{1 - (\norm{\svec_i - \cvec_k}/r)^2 \}^2 ~ \text{if} ~ \norm{\svec_i - \cvec_k} \leq r $ and $b_k(\svec_i) = 0 ~ \text{otherwise}$,
where $\cvec_k$ denotes the centroid of the basis function and $r$ specifying the range of the basis function. We select both of these hyperparameters using the standard options available in the \texttt{R} package \texttt{FRK} \citep{sainsbury2024modeling}, which distributes the centroids evenly across the observed spatial domain, while the number of basis functions $B$ is determined automatically by setting the resolution to be 2 or 3, resulting in up to a few hundred basis functions.
Turning to the weights for the basis functions $\alphavec$, we assumed these follow a multivariate normal distribution $\alphavec \sim \N_B(\zerovec, \Sigmavec)$, where the covariance matrix is constructed via an exponential covariance function i.e., $\Sigma_{k_1, k_2} = \sigma_{\alpha}^2 \exp\{-l_{\alpha}^{-1} \norm{\cvec_{k_1} - \cvec_{k_2}} \}; k_1, k_2 = 1,\dots,B$, where $\norm{\cvec_{k_1} - \cvec_{k_2}}$ denotes the Euclidean distance between the centroids, and $\sigma_{\alpha}$ and $l_{\alpha}$ are variance and scale parameters of the covariance function which are estimated during model fitting. Finally, $\xi(\svec_i)$ denotes a fine-scale variation term, and is assumed to follow a normal distribution with a common variance, $\xi(\svec_i) \sim \N(0, \sigma^2_{\xi}), i = 1,\dots, N$. 

\subsection{Estimation and prediction} \label{subsec:estimationprediction}

We employ maximum likelihood estimation to fit the siISDM as well as the existing models reviewed in Section \ref{subsec:existing_models}. Since the estimation procedure is largely similar across all models, we provide details for the proposed siISDM below, and only highlight differences in the case of other models as appropriate; details of the the marginal log-likelihood functions for the three models reviewed in Section \ref{subsec:existing_models} are provided in  Supplementary Material S2.1. 

Given fixed rank kriging is used to characterize the spatial random effect $u(\cdot)$ in \eqref{eq:single_index_eta}, and assuming the responses $y_{j}(\svec_i)$ for $i = 1, \dots, n_j; j = 1, \dots, m$ are conditionally independent given $\alphavec$ and $\xivec$, then the marginal log-likelihood function of the siISDM is given by
\begin{align*}
\ell_{\text{siISDM}}(\thetavec) = \log \left\{\int \prod_{j=1}^{m} \prod_{i=1}^{n_j} f\{y_{j}(\svec_i) \mid \mu_j(\svec_i), \phivec_j\} f(\alphavec \mid \sigma_{\alpha}^2, l_{\alpha}) f(\xivec \mid \sigma_{\xi}^2) \; \txd \alphavec \txd \xivec \right\},
\end{align*}
where $f\{y_{j}(\svec_i) \mid \mu_j(\svec_i), \phivec_j\}$ denotes the assumed distribution of the response e.g., the negative binomial distribution in the case of the scallop abundance data, $f(\alphavec \mid \sigma_{\alpha}^2, l_{\alpha}) = \N_B(\zerovec, \Sigmavec)$ with $\Sigmavec$ parametrized by $\sigma_{\alpha}^2$ and $l_{\alpha}$, and $f(\xivec \mid \sigma_{\xi}^2) = \N_N(\zerovec, \sigma^2_{\xi}\bm{I}_N)$ with $\bm{I}_N$ denoting an $N \times N$ identity matrix. The full vector of parameters characterizing the siISDM is given by $\thetavec = (\betavec^{\top}, \bm{\phi}_1^\top, \ldots, \bm{\phi}_m^\top, \sigma_{\alpha}^2, l_{\alpha}, \sigma_{\xi}^2, \bm{\psi})^{\top}$, where $\bm{\psi}$ generically denotes the parameters characterizing the CEFs e.g., $\bm{\psi} = (\gammavec_1^{\top}, \dots, \gammavec_m^{\top})^\top$ in the case of the four-parameter logistic function in \eqref{eq:logistic_link_fn}. 

Since $\ell_{\text{siISDM}}(\thetavec)$ does not in general possess a tractable form,
we employ a Laplace approximation and couple it with automatic differentiation techniques to efficiently maximize the Laplace approximated marginal likelihood function. Specifically, we use the \texttt{R} package \texttt{TMB} \citep{osgood2023statistical} 
to accomplish this; we obtain maximum likelihood estimates, $\hat \thetavec$, predictions of the basis function weights, $\hat \alphavec$, along with the corresponding joint observed Fisher information matrix $\hat{\mathcal{I}}(\hat \thetavec, \hat \alphavec)$, and the estimated asymptotic covariance matrix of the fixed-effects parameters given by $\hat{\mathcal{I}}(\hat \thetavec)^{-1}$. Standard errors and statistical inference for the effects of environmental and habitat factors on the index of true species abundance distribution, $\hat \betavec$, along with prediction of the associated spatial random effect at one or more spatial locations $\hat{u}(\svec_i) = \bvec(\svec_i)^\top \hat \alphavec$, then follow based on the above quantities. 

In the context of the siISDM, there are arguably two main quantities of ecological interest we want to construct predictions for: the index representing the true species abundance distribution in \eqref{eq:single_index_eta}, which is common across all data sources, and the survey-specific mean abundances and the associated responses based on \eqref{eq:mean_indexmodel} i.e., $\mu_j(\svec_i) = \exp\{\eta_j(\svec_i)\}$ and $y_j(\svec_i) \sim f\{y_j(\svec_i) \mid \mu_j(\svec_i), \bm{\phi}_j\}$. Predicting the former offers insight into how the underlying true species distribution is distributed across environmental and geographic space, while predicting the latter is important for understanding both how this is mapped to the observed abundance given a specific survey's characteristics, and to assess prediction performance in general.
To achieve both predictions, we adopt a simulation-based approach which accounts for the sampling errors of the model parameters as follows \citep[see also][]{fletcher2022single}: for $t = 1,\ldots,T$: 
1) sample parameter estimates and basis function weights from the estimated large sample distribution, $(\thetavec^{(t) \top}, \alphavec^{(t) \top} )^\top \approx \N\{(\hat \thetavec^\top, \hat \alphavec^\top)^\top, \hat{\mathcal{I}}(\hat \thetavec, \hat \alphavec)^{-1}\}$, as well as the fine scale variation term $\xivec^{(t)} \approx \N(\zerovec, \hat{\sigma}_{\xi}^{2} \Ivec)$; 
2) construct $\kappa^{(t)}(\svec^*)$ from $\thetavec^{(t)}, \alphavec^{(t)} , \xivec^{(t)}$ based on \eqref{eq:single_index_eta};
3) sample $y_j^{(t)}(\svec^*)$ from the distribution $f\{y_{j}(\svec^*) \mid \mu_j^{(t)}(\svec^*), \bm{\phi}^{(t)}_j\}$, where $\mu_j^{(t)}(\svec^*)$ is constructed based on \eqref{eq:mean_indexmodel} and the samples from steps 1 and 2. By repeating the above process a large number of times (we used $T = 500$ throughout this article), we obtain a sample of predictions $\{ y_j^{(1)}(\svec^*), \dots, y_j^{(T)}(\svec^*) \}$.
To avoid outliers which can arise during the simulation process, especially for overdispersed count responses, we can also consider trimming the predictions following a similar idea to that of \citet{cantoni2017random} e.g., we trim 2.5\% of the predictions $\{ y_j^{(1)}(\svec^*), \dots, y_j^{(T)}(\svec^*) \}$ from either tail prior to constructing point predictions and prediction intervals.

\section{Numerical study} \label{sec:simulation}

We performed a simulation study to assess the proposed siISDM in terms of estimating covariate effects and CEFs, and compare its predictive performance of survey-specific recorded abundance with independent SDMs and the additive field ISDMs.
We do not consider the additive constant ISDMs in the simulation due to its anticipated poor performance given our data generation process will involve non-constant differences in CEFs.

We generate species abundance from $m = 2$ surveys using an siISDM as follows. First, we set the spatial locations to be on an equally spaced $51 \times 51$ grid in the unit square $[0,1]^2$, and simulate a spatial field via the fixed rank kriging approach discussed towards the end of Section \ref{sec:single_index_model}, with centroids placed in the domain $[0,1] \times [0,1]$ (see Supplementary Material S3 
for a visualization of this) and $\alphavec$ generated from a multivariate normal distribution with an exponential covariance function with $\sigma^2_\alpha = 1$ and $l_\alpha = 0.1$. Note the true data generation process does not include a fine-scale term $\xi(\cdot)$. Next, across all spatial locations, we simulated a vector of four covariates $\xvec(\svec_i) = \{x_1(\svec_i), x_2(\svec_i), x_3(\svec_i), x_4(\svec_i)\}^{\top}$ such that covariates 1 to 3 were each generated from a Gaussian process with exponential covariance functions, where the variance in all three was set to 0.1 while the scale was set to 0.05, 0.02 and 0.01, respectively. The fourth covariate was simulated independently from a uniform distribution $[0,1]$. Note the data generation process is such the spatial field is smoother than the covariates, so as to minimize potential concerns around spatial confounding \citep{paciorek2010importance,hui2024spatial}. 
An index of the true species abundance distribution is then constructed using the above variables and \eqref{eq:single_index_eta}, where we set the vector of true covariate effects to be $\betavec = (1, 0.5, -1, -0.5)^{\top}$.
Finally, given the single index we then simulated count responses for the two surveys using \eqref{eq:mean_indexmodel}, and considering the following three pairs of CEFs:
\begin{itemize}
\item \underline{Scenario 1}: $h_1(\eta) = 5\{1 + \exp(-\eta)\}^{-1}$ and $h_2(\eta) = 5\{1 + \exp(-\eta)\}^{-1} + 2$. Here, both surveys share a common shape for the CEF, but the second survey is scaled up by a constant relative to the first survey;

\item \underline{Scenario 2}: $h_1(\eta) = 5\{1 + \exp(-\eta + 1)\}^{-1} - 1$, and $h_2(\eta) = 5\{1 + \exp(-\eta - 2)\}^{-1} - 1$. The first survey's CEF is relatively similar to that in Scenario 1. However, the second survey's CEF rises more quickly as a function of the index of true species distribution before plateauing. 
Both functions plateau to the same value 
e.g., sets of gear characterizing the two surveys exhibit the same limiting capacity;

\item \underline{Scenario 3}: $h_1(\eta) = 6\{1 + \exp(-\eta + 1)\}^{-1} - 1$, and $h_2(\eta) = 4\{1 + \exp(-\eta)\}^{-1}$. This is similar to Scenario 2 except the first survey's CEF crosses the second survey's and reaches a higher value as the index of true species distribution becomes larger. Ecologically, this could mean that the gear characterizing the first survey has a higher capacity limit i.e., is saturated at relatively higher true species abundance.
\end{itemize}

\begin{figure}[t!]
\centering
\includegraphics[width = \textwidth]{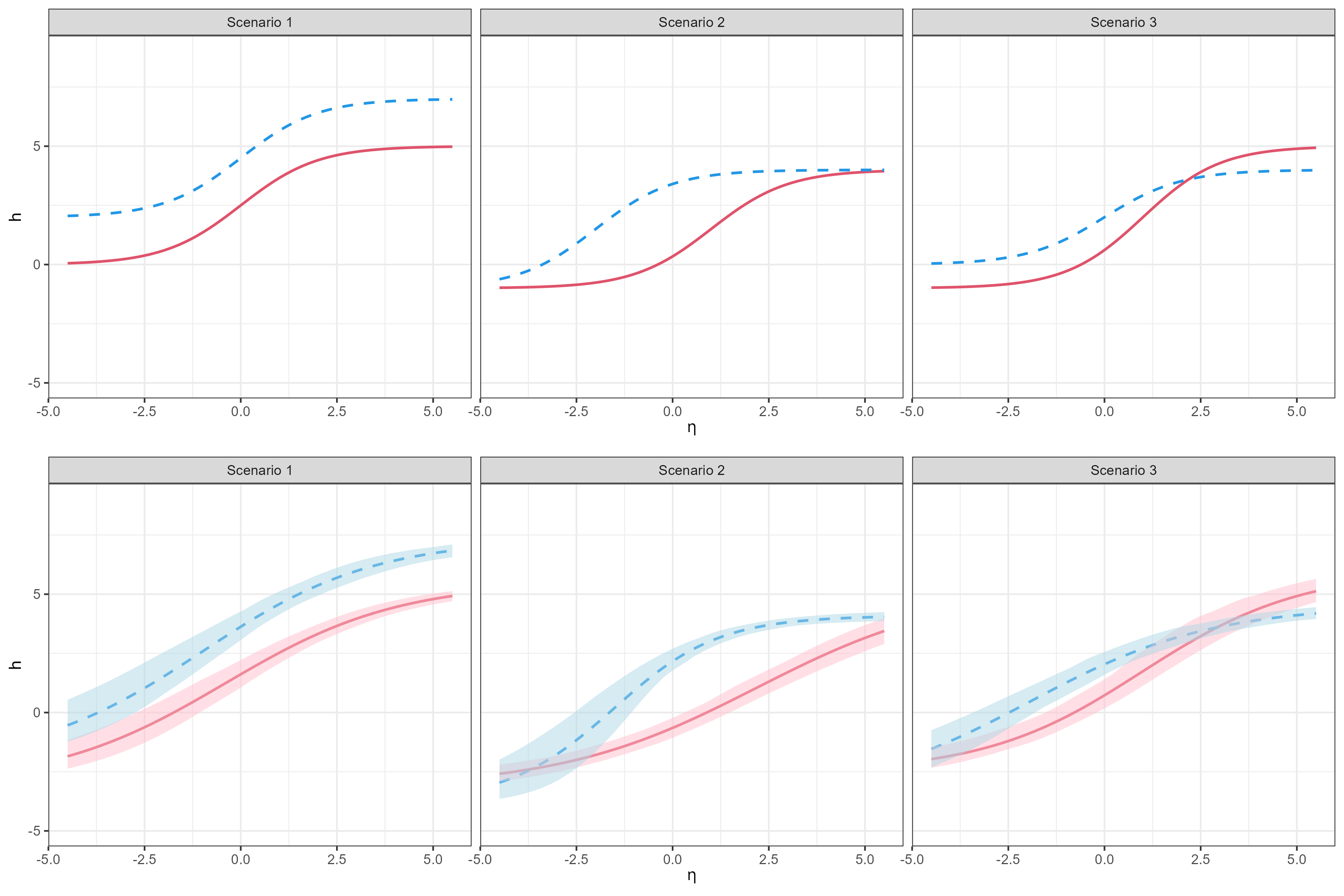}
\caption{True (top) and estimated (bottom) CEFs for the simulation study. Each column represents one of the three CEF scenarios. 
The solid red/blue dashed lines correspond to functions  $h_1(\cdot)$ and $h_2(\cdot)$ respectively. In the bottom row, the average estimated CEF across 100 simulated datasets is presented, while the shaded region corresponds to 95\% pointwise intervals.
}
\label{fig:sim_link_fn}
\end{figure}

All three pairs of true CEFs are visualized in the top row of Figure \ref{fig:sim_link_fn}. For each scenario, we simulated responses $y_j(\bm{s}_i)$ from a negative binomial distribution with survey-specific dispersion parameters $\phi_1 = 1$ and $\phi_2 = 2$.

Next, to reflect how fisheries abundance data may be sampled under different protocols in practice, and also to assess predictive performance, we allow the surveys to be sampled from different regions; see also the discussion around Figure \ref{fig:obs}. 
Specifically, for the first survey we randomly sampled $n_1 = 2,081$ observations from the entire spatial domain $[0,1]^2$, corresponding to 80\% of all spatial locations in the grid.
For the second survey, we divided the spatial domain into two spatial regions $[0,1] \times [0,0.5]$ and $[0,1] \times [0.5,1]$, and randomly sampled $n_2 = 1,061$ observations from the first spatial region only i.e., corresponding to 80\% of all spatial locations available in first subregion. Note this means observations in the latter subregion i.e., $[0,1] \times [0.5,1]$, are missing from all samples. Ecologically, we can view this as the first survey having a wider spatial coverage than the second survey. An example of the subsampled data across the two surveys presented in Supplementary Material S3. 
All models are fitted to the subsampled data to obtain the parameter estimates, while predictive performance is assessed on the non-sampled locations.
We repeated the above simulation 100 times i.e., generated 100 datasets and subsampling, for each of the CEF scenario above.

\subsection{Simulation results}
\begin{figure}[t]
\centering
\includegraphics[width = 0.8\textwidth]{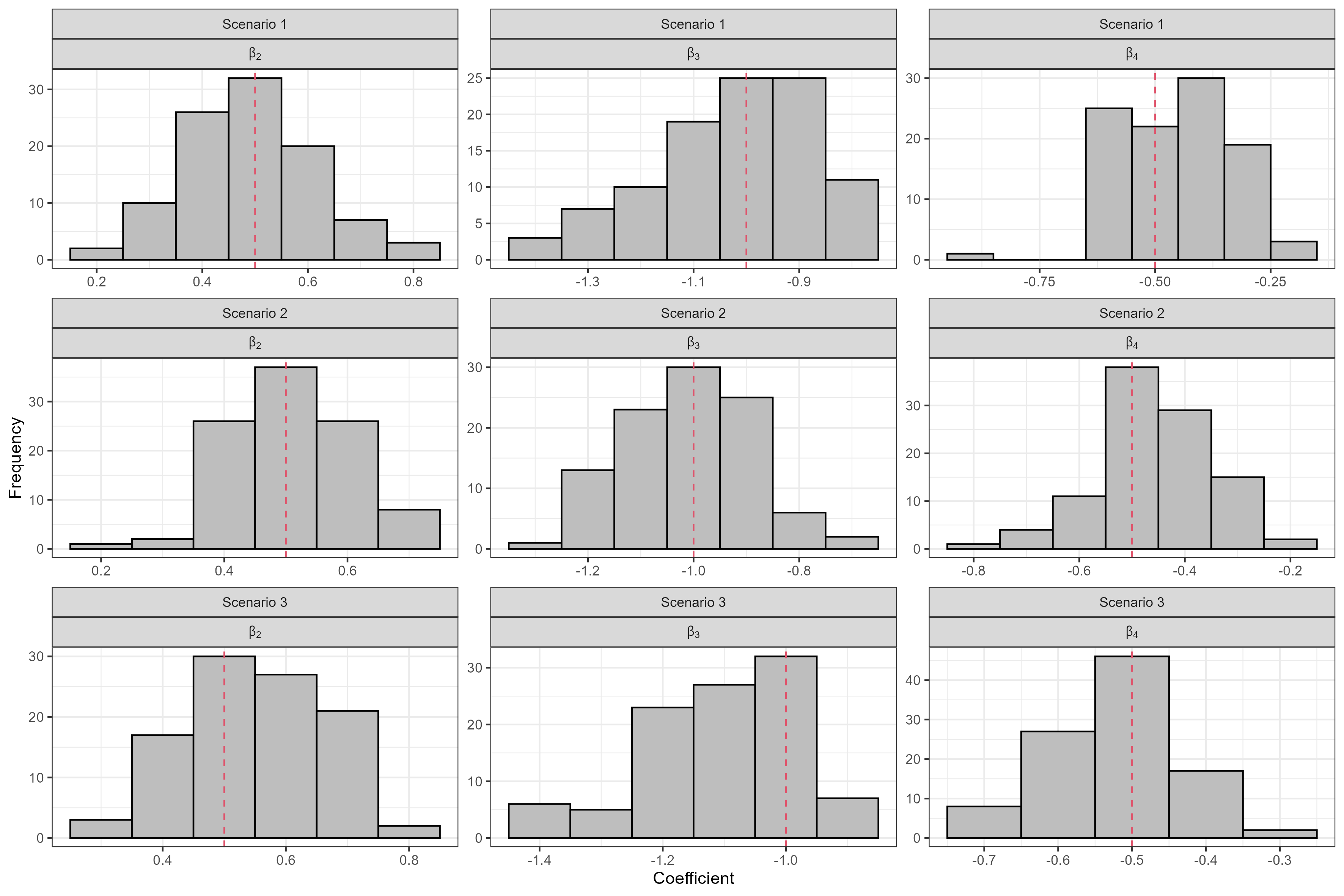}
\caption{Histograms of estimated covariate effects, obtained from fitting the siISDM to 100 simulated datasets in Scenarios 1 to 3 (top to bottom rows, respectively) of the simulation study. The three columns correspond to second, third and fourth coefficients in $\betavec$. The red vertical dashed line denote true parameter values.
}
\label{fig:sim_beta}
\end{figure}

We divide the presentation of our results into two parts. First, we fitted the siISDM proposed in Section \ref{sec:single_index_model} with the four-parameter logistic function \eqref{eq:logistic_link_fn}, and assessed our capacity to recover the covariate effects $\betavec$ and the CEFs $h_j(\cdot)$.
Figure \ref{fig:sim_beta} presents the histograms for the estimated covariate effects for each of the three scenarios, noting that the first coefficient is fixed to $1$ for identifiability. Overall, the coefficients $\hat \betavec$ appear to be reasonably well estimated with relatively minimal bias
for all scenarios.
The bottom row of Figure \ref{fig:sim_link_fn} presents the corresponding estimated CEFs, which closely resemble in shape and scale the corresponding true CEFs, albeit being slightly smoother, across all three scenarios.

Next, we compared the predictive performance of the siISDM with independent SDMs and additive field ISDMs detailed in Section \ref{subsec:existing_models}. Specifically, in each simulated dataset we fitted each model to the subsampled data described at the end of the previous section, and then predicted survey-specific recorded abundances $y_j(\svec_i); j = 1,2; i = 1,\ldots,n_{j,\text{test}}$ where $n_{j,\text{test}}$ denotes the number of non-sampled (test) observations for survey $j$. We assessed performance using four metrics: (i) root mean squared error of prediction or RMSPE; (ii) continuous rank probability score or CRPS \citep{gneiting2007strictly}; (iii) scaled CRPS or SCRPS \citep{bolin2023local} which also takes into account the scale of the prediction, and; (iv) interval score or IS based on 95\% prediction intervals. 
Figure S6 
in the Supplementary Material presents comparative boxplots for these prediction metrics. Focusing on the second survey, which has missing data for an entire subregion, we see that siISDMs produced the best performance across the three scenarios of CEFs. 
Interestingly, in both Scenarios 2 and 3 the additive field ISDM sometimes performed worse than the independent SDM: this is partly due to the fact the CEFs in these two scenarios are relatively more complicated (they are not just a location shift), 
and so the additive field ISDM struggled to attribute catch efficiency differences solely to differences in spatial random effects between the two surveys.

In Supplementary Material S3,
we present an additional simulation design where the true data generating process is an additive field ISDM, focusing solely on prediction in this setting. Overall, results for this show that 
the siISDM performed only slightly worse than the additive field ISDMs across a number of metrics, and particularly when the differences in spatial random effects between the two surveys were slight.

\section{Application to scallop abundance data} \label{sec:application_results}

We applied the siISDM, using both the four-parameter logistic function and the spline-based approach to model the CEFs, to the motivating scallop abundance data detailed in Section \ref{sec:data}. For the latter, we set the degree and the total number of I-spline basis functions to be $k = 2$  as $V = 7$, respectively; see also Supplementary Material S2.2 for detailed explanation of the spline construction . 
As environmental and habitat factors, we included both linear and quadratic terms for the nine variables detailed in Section \ref{sec:data} in the index of true species abundance distribution in \eqref{eq:single_index_eta}. 

We present the table of estimated covariate effects $\hat \betavec$ along with corresponding standard errors in Table \ref{tbl:beta_coeff}. 
Recalling these effects have to be interpreted in a relative sense, we observe that the index of underlying true species distribution is most strongly affected by bottom temperature 
($\hat \beta_{\text{Temp}} = -0.88$ and $\hat \beta_{\text{I(Temp$^2$)}}$ = -0.36 for the four-parameter logistic siISDM, and $\hat \beta_{\text{Temp}} =$ -0.85 and $\hat \beta_{\text{I(Temp$^2$)}} = -0.37$ for the spline-based siISDM), with a decreasing curve over the observed temperature range, generally consistent with known physiological tolerances;
and grain size ($\hat \beta_{\text{Grain}} = -0.52$ and $\hat \beta_{\text{I(Grain$^2$)}}$ = 0.19 for the four-parameter logistic siISDM, and $\hat \beta_{\text{Grain}} = -0.45$ and $\hat \beta_{\text{I(Grain$^2$)}} = 0.17$
for the spline-based siISDM), again with a decreasing curve, consistent with the idea adult scallops tend to be associated with sandier or more gravelly sediments. Most covariate effects in the index of true species distribution are statistically significant at the 5\% level, and moreover
both approaches to estimating the CEFs led to similar conclusions on the relative signs and magnitudes of the covariate effects on the index. Both siISDMs also produced similar survey-specific dispersion parameters for the negative binomial distribution: $(\hat \phi_1, \hat \phi_2, \hat \phi_3)^\top = (5.11, 4.97, 1.39)^\top$ for the four-parameter logistic function siISDM, and $(\hat \phi_1, \hat \phi_2, \hat \phi_3)^\top = (4.31, 4.41, 1.24)^\top$ for the semiparametric spline-based siISDM. Substantially larger overdispersion parameters are estimated for the two trawl surveys: this is expected given the substantially large number of zero counts recorded in these two surveys; see also Figure \ref{fig:obs}. 

\begin{table}[h!]
	\centering
	\caption{Estimated coefficients $\betavec$ and their standard errors for the single index ISDMs applied to the scallop abundance data, with the four-parameter logistic (4PL) catch efficiency functions and the spline-based catch efficiency functions. Asterisk indicates significance at 5\%.
    }
	\label{tbl:beta_coeff}
	\bgroup
	\begin{tabular}{lcc}
	\toprule[1.5pt]
	& \multicolumn{1}{c}{4PL siISDM}
	& \multicolumn{1}{c}{Spline-based siISDM} \\
	\midrule
Salinity                & {1.00 (-)}   & {1.00 (-)}   \\
I(Salinity\textsuperscript{2}) & {0.28 (0.03)*}  & {0.30 (0.02)*}  \\
Temp                 & {-0.88 (0.03)*} & {-0.85 (0.02)*} \\
I(Temp\textsuperscript{2}) & {-0.36 (0.02)*} & {-0.37 (0.02)*} \\
Stress                           & {0.01 (0.05)} & {-0.02 (0.04)} \\
I(Stress\textsuperscript{2})     & {-0.35 (0.03)*} & {-0.26 (0.02)*} \\
Depth                                  & {-1.14 (0.06)*} & {-1.25 (0.05)*} \\
I(Depth\textsuperscript{2})            & {-1.73 (0.05)*} & {-1.80 (0.04)*} \\
Velocity                             & {-0.39 (0.04)*} & {-0.42 (0.04)*} \\
I(Velocity\textsuperscript{2})       & {0.03 (0.02)}  & {0.02 (0.02)}  \\
CHL                         & {-0.13 (0.03)*} & {-0.12 (0.03)*} \\
I(CHL\textsuperscript{2})    & {-0.36 (0.03)*} & {-0.33 (0.02)*} \\
Complexity                                & {-0.02 (0.02)} & {-0.02 (0.02)}  \\
I(Complexity\textsuperscript{2})          & {-0.06 (0.02)*} & {-0.05 (0.01)*} \\
BPI                                    & {-0.02 (0.02)}  & {-0.05 (0.02)}  \\
I(BPI\textsuperscript{2})              & {-0.24 (0.02)*} & {-0.23 (0.02)*} \\
Grain                                     & {-0.52 (0.03)*} & {-0.45 (0.01)*} \\
I(Grain\textsuperscript{2})               & {0.19 (0.01)*}  & {0.17 (0.01)*}  \\
	\bottomrule[1.5pt]
	\end{tabular}
	\egroup
\end{table}

\begin{figure}[t!]
\centering
\includegraphics[width = 0.8\textwidth]{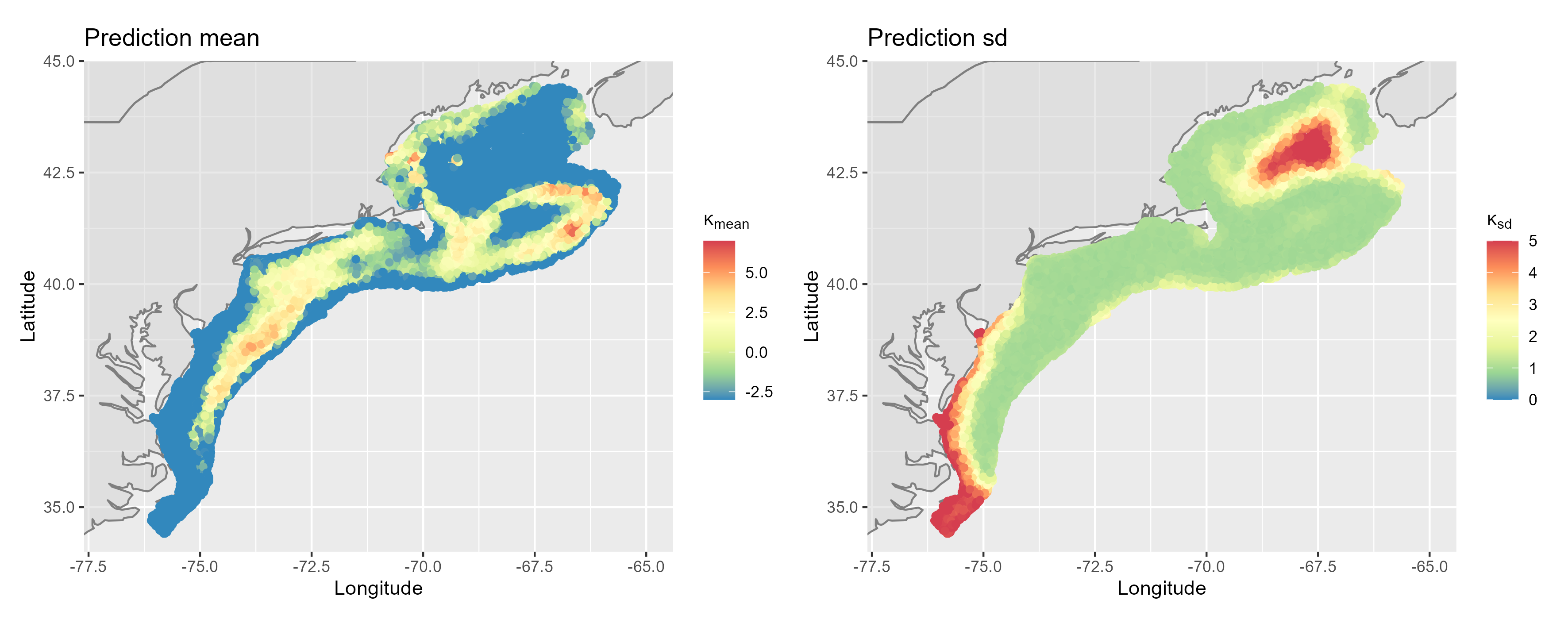}
\includegraphics[width = 0.8\textwidth]{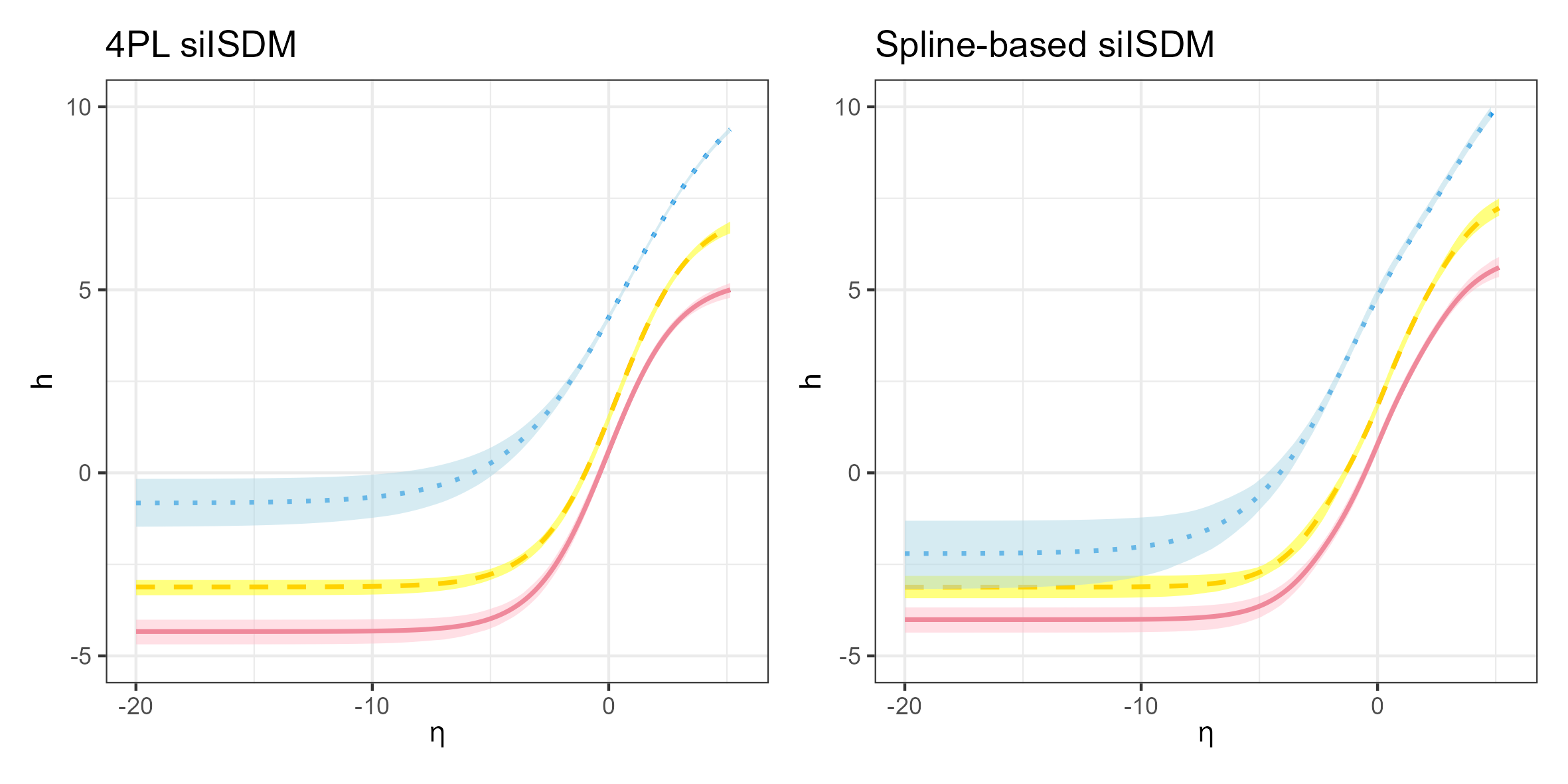}
\caption{Top row: Estimated single index $\kappa(\cdot)$ for the four-parameter logistic siISDM, with the predicted mean (left)  standard deviation of $\kappa(\cdot)$ (right) shown. 
Bottom row: Estimated CEFs in the four-parameter logistic function (left) and semiparametric spline-based (right) siISDM applied to the scallop abundance data. In both panels, the solid red line corresponds to $h_1(\cdot)$ for the pre-2009 bottom trawl survey, the dashed orange line corresponds to $h_2(\cdot)$ for the post-2009 bottom trawl survey, while the blue dotted line corresponds to the $h_3(\cdot)$ for the scallop dredge survey. 
}
\label{fig:scallop_kappa_CEFs}
\end{figure}

Next, the top row of Figure \ref{fig:scallop_kappa_CEFs} presents the estimated index of true species abundance i.e., $\kappa(\cdot)$, obtained using the four-parameter logistic function siSIDM; the same plots for the semiparametric spline-based siISDM are similar and so omitted for brevity. The siISDM identified particular spatial subregions between 41.0$\degree$N-42.5$\degree$N and 67.5$\degree$W-66.0$\degree$W, and between 38.0$\degree$N-39.0$\degree$N and 74.0$\degree$W-73.0$\degree$W, as having the highest abundance of scallops. By contrast, the coastal region from 35.0$\degree$N-40.0$\degree$N and the spatial region between 41.5$\degree$N-44.5$\degree$N and 70.0$\degree$W-66.0$\degree$W are expected to have few to no scallops present.
The bottom row of Figure \ref{fig:scallop_kappa_CEFs} presents the estimated CEFs produced by both the four-parameter logistic and the spline-based siISDMs. The overall shapes estimated using both approaches are fairly similar. In particular, the CEF of the survey $h_2(\cdot)$ is effectively a positive location shifted version of the first survey's function $h_1(\cdot)$; see also Scenario 1 in the simulations in Section \ref{sec:simulation}. This is sensible given the former corresponds to the post-2009 bottom trawl survey, which as explained in Section \ref{sec:data} used more modern vessels and gears with improved bottom contact. Thus while the shape of the mapping from true species abundance to recorded abundance is expected to remain similar, an overall increase in catch efficiency is anticipated; the siISDM was able to capture this explicitly in the estimated CEFs. 
The estimated CEF for the scallop dredge survey $h_3(\cdot)$ is notably larger than both $h_1(\cdot)$ and $h_2(\cdot)$. This was again consistent with the targeted sampling protocol and specialized gear design of this dredge survey. 
Moreover, the shape of the dredge CEF also differed slightly from that of the trawls, as it begins to increase at lower levels of underlying abundance, suggesting it is capable of detecting density variation at lower levels. Furthermore, the dredge CEF was yet to plateau towards large (observed) values of the index of true species distribution, suggesting that it also better detects differences at high underlying scallop densities.

Finally, we assessed the predictive performance of the siISDM, comparing it to independent species distribution, additive field ISDMs and additive constant ISDMs; note all but the additive constant ISDM were also considered in the simulation study in Section \ref{sec:simulation}. We perform a five-fold spatial block cross-validation \citep{valavi2019block} to assess predictive performance, where the spatial locations for each survey were first divided into spatial blocks and assigned numbers 1 to 5. Then in the $k$th fold, we leave out the blocks labeled $k$ and fit the models to all remaining blocks across all three surveys. We assessed performance using the same four metrics as in Section \ref{sec:simulation}. Overall, Table \ref{tbl:block_cross_valid} shows that the four-parameter logistic function siISDM performed best across the three surveys, followed closely by the semiparametric spline-based siISDM. Perhaps unsurprisingly, independent SDMs performed the poorest, with the most dramatic case being the scallop dredge survey. Interestingly, the additive field ISDM was not always better than the additive constant ISDM, e.g., for the data-rich scallop dredge survey the additive field ISDM actually performed worse than its additive constant counterpart across all four metrics.
Such a result is likely due the spatial random effects in the additive field ISDM potentially overfitting the data, while a simpler model such as the additive constant ISDM, as well as the siISDM (that more explicitly attempts to define the mapping between true species abundance and survey-specific recorded abundance) leads to better integration of multi-source data and thus prediction.

\begin{table}[t]
    \centering
    \caption{Average predictive metrics (RMSPE, CRPS, SCRPS, IS) for the three surveys based on applying five-fold spatial block cross-validation to the scallop abundance data. Four methods are compared: independent SDMs (Independent), additive constant ISDMs (Add. Constant), additive field ISDMs (Add. Field), and two flavors of siISDMs based on the four-parameter logistic function (4PL) and semiparametric spline-based approaches. Highlighted values indicate model with the best metric. }
    \label{tbl:block_cross_valid}
    \bgroup
    \scalebox{0.85}{
    \begin{tabular}{llrrrrr}
        \toprule[1.5pt]
        Survey & Metric & Independent & Add. Constant & Add. Field & 4PL siISDM & Spline-based siISDM \\
        \cmidrule{3-7}
            & RMSPE  & 158.2  & 149.2  & 352.7  & \textbf{141.8}  & 146.0 \\
            Pre-2009 & CRPS   & 14.6   & 14.5   & 16.6   & \textbf{14.3}    & 14.9 \\
            trawl & SCRPS  & 11.4   & 46.0   & \textbf{5.9}    & 36.3   & 26.8 \\
            survey & IS     & 447.4  & 416.2  & 456.7  & \textbf{402.1}  & 425.0 \\\\
            & RMSPE  & 490.0  & 354.3  & 348.7  & \textbf{340.8}  & 341.4 \\
            Post-2009 & CRPS   & 43.7   & 39.3   & 39.0  & \textbf{38.8}   & 39.3 \\
            trawl & SCRPS  & \textbf{4.7}    & 9.5    & 7.7    & 9.9   & 12.7 \\
            survey & IS     & 1490.5 & 1086.9 & 1064.0 & 1042.1 & \textbf{1024.3} \\\\
            & RMSPE  & 2716.3 & 1136.8 & 1362.5 & \textbf{1105.9} & 1180.2 \\
            Scallop & CRPS   & 527.1  & 364.0  & 396.3  & \textbf{358.3}  & 368.2 \\
            dredge & SCRPS  & 32.0   & 9.2    & 19.0   & 8.2    & \textbf{7.5} \\
            survey & IS     & 13872.3 & 9331.9 & 10067.1 & \textbf{8871.4} & 9068.0 \\
        \bottomrule[1.5pt]
    \end{tabular}
    }
    \egroup
\end{table}

\section{Discussion} \label{sec:discussion}

In this article, motivated by the collection of different surveys of the same species in fisheries ecology, we propose a single index integrated species distribution model to combine data from multiple sources that have distinct characteristics. The siISDM uses a single index to represent the underlying true species abundance distribution, and combines this with survey-specific functions to represent the catch efficiency properties of different surveys used to collect the species. 
In doing so, siISDM is arguably more aligned with the ecological and sampling processes driving the observed patterns of species abundance across heterogeneous surveys.
Through simulation studies and an analysis of scallop abundance records collected by two bottom trawl and one dredge survey, we show that our proposed siISDM can provide meaningful interpretation for the covariate effects, the underlying species abundance, and the difference in catchabilities of the surveys.
Additionally, the siISDM can offer strong predictive performance compared to the commonly used additive field ISDM as well as naive independent SDMs fitted to each survey.


There are a number of further developments that can be made for siISDMs. 
For instance, we currently assume the random effects to be a spatial field, but a natural extension is to make the random effects spatio-temporal if the surveys are spatially- and temporally-indexed, although an additional challenge here is the subsequent increase in computational complexity. 
As the spatial field is observed over the ocean, more complicated construction of the spatial field, such as one accounting for physical barriers \citep{bakka2019non} might also be used.
Also, the siISDM combines a data-driven component (represented by the spatial field) with a science-informed component (represented by the survey-specific CEFs). In our application to the scallop abundance data, this approach outperforms the additive field ISDM which relies solely on the spatial fields to describe differences between surveys. This more broadly suggests that incorporating further scientific knowledge into ISDMs can potentially further enhance its performance. As an example, for some species, especially those surveyed with more complex gear or those exhibiting complex interactions with their habitats,  it may be appropriate to allow the parameters $\gammavec$ characterizing the CEFs $h_j(\cdot)$ to vary with environmental predictors and/or space \citep[a type of spatially varying single index model e.g.,][]{luo2016single}.
Equally, we acknowledge that survey-specific discrepancies in catch efficiency can arise through a wide variety of mechanisms beyond differences in fishing gear type. For example, when spatially-varying processes variability in both substrate type and complexity are the principal drivers of catch efficiency differences \citep{delargy2023global}, then there may be much less practical difference between the proposed siISDM and the more flexible additive field ISDM.

Finally, beyond fisheries abundance in ecology, the idea of data integration has become a popular topic in the statistics literature of late \citep[see the recent reviews of][]{han2025frontiers}. 
We believe the single index approach may provide an alternative to, but also learn from, recent developments in this literature e.g., using method-of-moment techniques for more scalable computation or using hierarchical/fusion-type penalties to borrow strength for the CEFs $h_j(\cdot)$ \citep[e.g.,][]{hector2021distributed,hui2024homogeneity}.

\section*{Acknowledgements}
This work was supported by the Australian Research Council under Grants DP230101908 and DP240100143.

\bibliographystyle{apalike}
\bibliography{biblio}

@article{hui2024homogeneity,
  title={{Homogeneity pursuit and variable selection in regression models for multivariate abundance data}},
  author={Hui, Francis K C and Maestrini, Luca and Welsh, Alan H},
  journal={Biometrics},
  volume={80},
  pages={ujad001},
  year={2024},
  publisher={Oxford University Press}
}

@article{han2025frontiers,
  title={{Frontiers in data integration}},
  author={Han, Peisong and Si, Yajuan},
  journal={Journal of the Royal Statistical Society: Series A},
  volume={188},
  pages={24--26},
  year={2025},
  publisher={Oxford University Press UK}
}

@article{hector2021distributed,
  title={{A distributed and integrated method of moments for high-dimensional correlated data analysis}},
  author={Hector, Emily C and Song, Peter X-K},
  journal={Journal of the American Statistical Association},
  volume={116},
  pages={805--818},
  year={2021},
  publisher={Taylor \& Francis}
}

@article{paciorek2010importance,
  title={{The importance of scale for spatial-confounding bias and precision of spatial regression estimators}},
  author={Paciorek, Christopher J},
  journal={Statistical Science},
  volume={25},
  pages={107--125},
  year={2010}
}

@article{osgood2023statistical,
  title={{A statistical review of template model builder: A flexible tool for spatial modelling}},
  author={Osgood-Zimmerman, Aaron and Wakefield, Jon},
  journal={International Statistical Review},
  volume={91},
  pages={318--342},
  year={2023},
  publisher={Wiley Online Library}
}

@article{dumbgen2024shape,
  title={{Shape-constrained statistical inference}},
  author={D{\"u}mbgen, Lutz},
  journal={Annual Review of Statistics and Its Application},
  volume={11},
  year={2024},
  publisher={Annual Reviews}
}

@article{wang2009spline,
  title={{Spline estimation of single-index models}},
  author={Wang, Li and Yang, Lijian},
  journal={Statistica Sinica},
  pages={765--783},
  volume = {19},
  year={2009},
  publisher={JSTOR}
}

@article{heaton2019case,
  title={{A case study competition among methods for analyzing large spatial data}},
  author={Heaton, Matthew J and Datta, Abhirup and Finley, Andrew O and Furrer, Reinhard and Guinness, Joseph and Guhaniyogi, Rajarshi and Gerber, Florian and Gramacy, Robert B and Hammerling, Dorit and Katzfuss, Matthias and Lindgren, F and Nychka, D W and Sun, F and Zammit-Mangion, A},
  journal={Journal of Agricultural, Biological and Environmental Statistics},
  volume={24},
  pages={398--425},
  year={2019},
  publisher={Springer}
}

@book{cressie2011statistics,
  title={{Statistics for Spatio-Temporal Data}},
  author={Cressie, Noel and Wikle, Christopher K},
  year={2011},
  publisher={John Wiley \& Sons}
}

@article{van2021model,
  title={{Model-based ordination for species with unequal niche widths}},
  author={van der Veen, Bert and Hui, Francis K C and Hovstad, Knut A and Solbu, Erik B and O'Hara, Robert B},
  journal={Methods in Ecology and Evolution},
  volume={12},
  pages={1288--1300},
  year={2021},
  publisher={Wiley Online Library}
}

@article{austin2002spatial,
  title={{Spatial prediction of species distribution: an interface between ecological theory and statistical modelling}},
  author={Austin, Mike P},
  journal={Ecological Modelling},
  volume={157},
  pages={101--118},
  year={2002},
  publisher={Elsevier}
}

@article{stoklosa2022overview,
  title={{An overview of modern applications of negative binomial modelling in ecology and biodiversity}},
  author={Stoklosa, Jakub and Blakey, Rachel V and Hui, Francis K C},
  journal={Diversity},
  volume={14},
  pages={320},
  year={2022},
  publisher={MDPI}
}

@article{fletcher2019practical,
  title={A practical guide for combining data to model species distributions},
  author={Fletcher, Robert J and Hefley, Trevor J and Robertson, Ellen P and Zuckerberg, Benjamin and McCleery, Robert A and Dorazio, Robert M},
  journal={Ecology},
  volume={100},
  pages={e02710},
  year={2019},
  publisher={Wiley Online Library}
}

@article{rufener2021bridging,
  title={Bridging the gap between commercial fisheries and survey data to model the spatiotemporal dynamics of marine species},
  author={Rufener, Marie-Christine and Kristensen, Kasper and Nielsen, J Rasmus and Bastardie, Francois},
  journal={Ecological Applications},
  volume={31},
  pages={e02453},
  year={2021},
  publisher={Wiley Online Library}
}

@article{dovers2024fast,
  title={A fast method for fitting integrated species distribution models},
  author={Dovers, Elliot and Popovic, Gordana C and Warton, David I},
  journal={Methods in Ecology and Evolution},
  volume={15},
  pages={191--203},
  year={2024},
  publisher={Wiley Online Library}
}

@article{simmonds2020more,
  title={{Is more data always better? A simulation study of benefits and limitations of integrated distribution models}},
  author={Simmonds, Emily G and Jarvis, Susan G and Henrys, Peter A and Isaac, Nick JB and O'Hara, Robert B},
  journal={Ecography},
  volume={43},
  pages={1413--1422},
  year={2020},
  publisher={Wiley Online Library}
}

@article{sainsbury2024modeling,
  title={{Modeling big, heterogeneous, non-Gaussian spatial and spatio-temporal data using FRK}},
  author={Sainsbury-Dale, Matthew and Zammit-Mangion, Andrew and Cressie, Noel},
  journal={Journal of Statistical Software},
  volume={108},
  pages={1--39},
  year={2024}
}

@article{valavi2019block,
  title={{blockCV: An R package for generating spatially or environmentally separated folds for k-fold cross-validation of species distribution models}},
  author={Valavi, Roozbeh and Elith, Jane and Lahoz-Monfort, Jos{\'e} J and Guillera-Arroita, Gurutzeta},
  journal={Methods in Ecology and Evolution},
  volume={10},
  pages={225--232},
  year={2019},
  publisher={Wiley Online Library}
}

@article{hardle1993optimal,
  title={Optimal smoothing in single-index models},
  author={Hardle, Wolfgang and Hall, Peter and Ichimura, Hidehiko},
  journal={Annals of Statistics},
  volume={21},
  pages={157--178},
  year={1993},
  publisher={Institute of Mathematical Statistics}
}

@article{ramsay1988monotone,
  title={Monotone regression splines in action},
  author={Ramsay, James O},
  journal={Statistical Science},
  volume={3},
  pages={425--441},
  year={1988},
  publisher={JSTOR}
}

@article{hui2024spatial,
  title={Spatial confounding in joint species distribution models},
  author={Hui, Francis K C and Vu, Quan and Hooten, Mevin B},
  journal={Methods in Ecology and Evolution},
  volume={15},
  pages={1906--1921},
  year={2024},
  publisher={Wiley Online Library}
}

@article{cantoni2017random,
  title={A random-effects hurdle model for predicting bycatch of endangered marine species},
  author={Cantoni, Eva and Mills Flemming, J and Welsh, AH},
journal = {Annals of Applied Statistics},
volume = {11},
pages = {2178 -- 2199},
  year={2017}
}

@article{foster2013variable,
  title={Variable selection in monotone single-index models via the adaptive {LASSO}},
  author={Foster, Jared C and Taylor, Jeremy MG and Nan, Bin},
  journal={Statistics in Medicine},
  volume={32},
  pages={3944--3954},
  year={2013},
  publisher={Wiley Online Library}
}

@article{luo2016single,
  title={Single-index varying coefficient model for functional responses},
  author={Luo, Xinchao and Zhu, Lixing and Zhu, Hongtu},
  journal={Biometrics},
  volume={72},
  pages={1275--1284},
  year={2016},
  publisher={Wiley Online Library}
}

@article{gruss2023spatially,
  title={Spatially varying catchability for integrating research survey data with other data sources: case studies involving observer samples, industry-cooperative surveys, and predators as samplers},
  author={Gr{\"u}ss, Arnaud and Thorson, James T and Anderson, Owen F and O’Driscoll, Richard L and Heller-Shipley, Madison and Goodman, Scott},
  journal={Canadian Journal of Fisheries and Aquatic Sciences},
  volume={80},
  pages={1595--1615},
  year={2023},
  publisher={Canadian Science Publishing 1840 Woodward Drive, Suite 1, Ottawa, ON K2C 0P7}
}

@article{cressie2008fixed,
  title={Fixed rank kriging for very large spatial data sets},
  author={Cressie, Noel and Johannesson, Gardar},
  journal={Journal of the Royal Statistical Society: Series B},
  volume={70},
  pages={209--226},
  year={2008},
  publisher={Wiley Online Library}
}

@techreport{politis2014northeast,
  title={{Northeast Fisheries Science Center bottom trawl survey protocols for the NOAA ship Henry B. Bigelow}},
  author={Politis, Philip J and Galbraith, John K and Kostovick, Paul and Brown, Russell W},
  institution={NOAA},
  year={2014}
}

@article{hart2006long,
  title={{Long-term dynamics of US Atlantic sea scallop Placopecten magellanicus populations}},
  author={Hart, Deborah R and Rago, Paul J},
  journal={North American Journal of Fisheries Management},
  volume={26},
  pages={490--501},
  year={2006},
  publisher={Oxford University Press Oxford, UK}
}

@article{lellouche2021copernicus,
  title={{The Copernicus global 1/12 oceanic and sea ice GLORYS12 reanalysis}},
  author={Lellouche, Jean-Michel and Eric, Greiner and Romain, Bourdall{\'e}-Badie and others},
  journal={Frontiers in Earth Science},
  volume={9},
  pages={698876},
  year={2021},
  publisher={Frontiers Media SA}
}

@article{wilkin2022data,
  title={{A data-assimilative model reanalysis of the US Mid Atlantic Bight and Gulf of Maine: Configuration and comparison to observations and global ocean models}},
  author={Wilkin, John and Levin, Julia and Moore, Andrew and Arango, Hernan and L{\'o}pez, Alexander and Hunter, Elias},
  journal={Progress in Oceanography},
  volume={209},
  pages={102919},
  year={2022},
  publisher={Elsevier}
}

@misc{nefsc201050th,
  title={50th Northeast Regional Stock Assessment Workshop (50th SAW): Assessment Summary Report.},
  author={NEFSC},
  year={2010},
  publisher={Northeast Fisheries Science Center Reference Document 10-09}
}

@misc{coastalrelief2023,
  title={{Coastal Relief Models (CRMs) [Data set]}},
  author={{NOAA National Centers for Environmental Information}},
  year={2023}
}

@techreport{dalyander2012documentation,
  title={{Documentation of the US Geological survey sea floor stress and sediment mobility database}},
  author={Dalyander, P Soupy and Butman, Bradford and Sherwood, Christopher R and Signell, Richard P},
  year={2012},
  institution={US Geological Survey}
}

@article{isaac2020data,
  title={Data integration for large-scale models of species distributions},
  author={Isaac, Nick JB and Jarzyna, Marta A and Keil, Petr and others}, 
  journal={Trends in Ecology \& Evolution},
  volume={35},
  pages={56--67},
  year={2020},
  publisher={Elsevier}
}

@misc{miller2010estimation,
  title={{Estimation of Albatross IV to Henry B. Bigelow calibration factors}},
  author={Miller, Timothy Jason and Das, Chhandita and Politis, Philip J and Miller, Alicia S and Lucey, Sean M and Legault, Christopher M and Brown, Russell W and Rago, Paul J},
  year={2010}
}

@article{feng2021sparse,
  title={Sparse single index models for multivariate responses},
  author={Feng, Yuan and Xiao, Luo and Chi, Eric C},
  journal={Journal of Computational and Graphical Statistics},
  volume={30},
  pages={115--124},
  year={2021},
  publisher={Taylor \& Francis}
}

@article{mclean2014functional,
  title={Functional generalized additive models},
  author={McLean, Mathew W and Hooker, Giles and Staicu, Ana-Maria and Scheipl, Fabian and Ruppert, David},
  journal={Journal of Computational and Graphical Statistics},
  volume={23},
  pages={249--269},
  year={2014},
  publisher={Taylor \& Francis}
}

@article{balabdaoui2019least,
  title={Least squares estimation in the monotone single index model},
  author={Balabdaoui, Fadoua and Durot, C{\'e}cile and Jankowski, Hanna},
  journal={Bernoulli},
  volume={25},
  pages={3276--3310},
  year={2019},
  publisher={JSTOR}
}

@article{sathyendranath2019ocean,
  title={{An ocean-colour time series for use in climate studies: the experience of the ocean-colour climate change initiative (OC-CCI)}},
  author={Sathyendranath, Shubha and Brewin, Robert JW and Brockmann, Carsten and others}, 
  journal={Sensors},
  volume={19},
  pages={4285},
  year={2019}
}

@article{szpilka2016improvements,
  title={{Improvements for the western North Atlantic, Caribbean and Gulf of Mexico ADCIRC tidal database (EC2015)}},
  author={Szpilka, Christine and Dresback, Kendra and Kolar, Randall and Feyen, Jesse and Wang, Jindong},
  journal={Journal of Marine Science and Engineering},
  volume={4},
  pages={72},
  year={2016},
  publisher={MDPI}
}

@article{lundblad2006benthic,
  title={{A benthic terrain classification scheme for American Samoa}},
  author={Lundblad, Emily R and Wright, Dawn J and Miller, Joyce and Larkin, Emily M and Rinehart, Ronald and Naar, David F and Donahue, Brian T and Anderson, S Miles and Battista, Tim},
  journal={Marine Geodesy},
  volume={29},
  pages={89--111},
  year={2006},
  publisher={Taylor \& Francis}
}

@misc{poppe2014usgs,
  title={{USGS East Coast sediment analysis: procedures, database, and GIS data. US Geological Survey Open-File Report 2005-1001}},
  author={Poppe, LJ and McMullen, KY and Williams, SJ and Paskevich, VF},
  year={2014}
}

@article{fletcher2022single,
  title={Single-fit bootstrapping: A simple alternative to the delta method},
  author={Fletcher, David and Jowett, Tim},
  journal={Methods in Ecology and Evolution},
  volume={13},
  pages={1358--1367},
  year={2022},
  publisher={Wiley Online Library}
}

@article{ratkowsky1986choosing,
  title={Choosing near-linear parameters in the four-parameter logistic model for radioligand and related assays},
  author={Ratkowsky, David A and Reedy, Terry J},
  journal={Biometrics},
  volume={42},
  pages={575--582},
  year={1986}
}

@article{delargy2023global,
  title={A global review of catch efficiencies of towed fishing gears targeting scallops},
  author={Delargy, Adam J and Blackadder, Lynda and Bloor, Isobel and McMinn, Carrie and Rudders, David B and Szostek, Claire L and Dobby, Helen and Kangas, Mervi and Stewart, Bryce D and Williams, James R and others},
  journal={Reviews in Fisheries Science \& Aquaculture},
  volume={31},
  pages={296--319},
  year={2023},
  publisher={Taylor \& Francis}
}

@article{gneiting2007strictly,
  title={Strictly proper scoring rules, prediction, and estimation},
  author={Gneiting, Tilmann and Raftery, Adrian E},
  journal={Journal of the American Statistical Association},
  volume={102},
  pages={359--378},
  year={2007},
  publisher={Taylor \& Francis}
}

@article{bolin2023local,
  title={Local scale invariance and robustness of proper scoring rules},
  author={Bolin, David and Wallin, Jonas},
  journal={Statistical Science},
  volume={38},
  pages={140--159},
  year={2023},
  publisher={Institute of Mathematical Statistics}
}

@article{bakka2019non,
  title={Non-stationary {G}aussian models with physical barriers},
  author={Bakka, Haakon and Vanhatalo, Jarno and Illian, Janine B and Simpson, Daniel and Rue, H{\aa}vard},
  journal={Spatial Statistics},
  volume={29},
  pages={268--288},
  year={2019},
  publisher={Elsevier}
}

\FloatBarrier

\newpage
\appendix
\def\thesection{S\arabic{section}}
\def\thefigure{S\arabic{figure}}
\def\thetable{S\arabic{table}}
\setcounter{table}{0}
\setcounter{figure}{0}

\section{Additional details and results for application to scallop abundance data} \label{sec:additionalinfo_scallop}

\FloatBarrier

\subsection{Additional details on covariates}\label{sec:additional_details_covariates}

We provide additional details on the covariates used in the scallop data analysis in this section.
Physicochemical parameters at sampled locations included annual mean bottom temperature and salinity (averaged over the 12 months that preceded a sampling event) obtained from the GLORYS \citep[years 2000-2007,][]{lellouche2021copernicus} and DOPPIO \citep[years 2007-2019,][]{wilkin2022data} oceanographic reanalyses.  Annual mean dissolved chlorophyll-a concentration (a proxy for primary productivity) at sampling sites was estimated from Ocean Colour Climate Change Initiative V6.0, 4km 5-day multi-sensor composites \citep{sathyendranath2019ocean}, averaging over the 12 months prior to sampling. 
Static hydrodynamic variables included the intensity of water movement near the seabed due to surface gravity waves and broad-scale circulatory currents, quantified via the 95th percentile of annual bottom stress obtained from the USGS seabed stress database \citep{dalyander2012documentation}, and maximum tidal current velocity magnitudes estimated using harmonic constituents from the ADCIRC EC2015 Tidal database \citep{szpilka2016improvements}. Bathymetric data (i.e., water depths) were obtained from the NOAA National Centers for Environmental Information 1-arc-second coastal relief model \citep{coastalrelief2023}, and subsequently used to derive additional benthic habitat variables including Bathymetric Position Index \citep[BPI,][]{lundblad2006benthic}, with an inner radius of 9 cells and outer radius of 90 cells) as well as an index of structural complexity, taken as the standard deviation of fine-scale BPI (with an inner radius of 3 cells and outer radius of 15 cells), within a $3 \times 3$ neighborhood. 
Finally, sediment grain size (on the phi scale) was estimated at each sample location based on observations from the U.S. Geological Survey East-Coast Sediment Texture Database \citep{poppe2014usgs}.

\begin{figure}[h!]
\centering
\includegraphics[width = 1.1\textwidth]{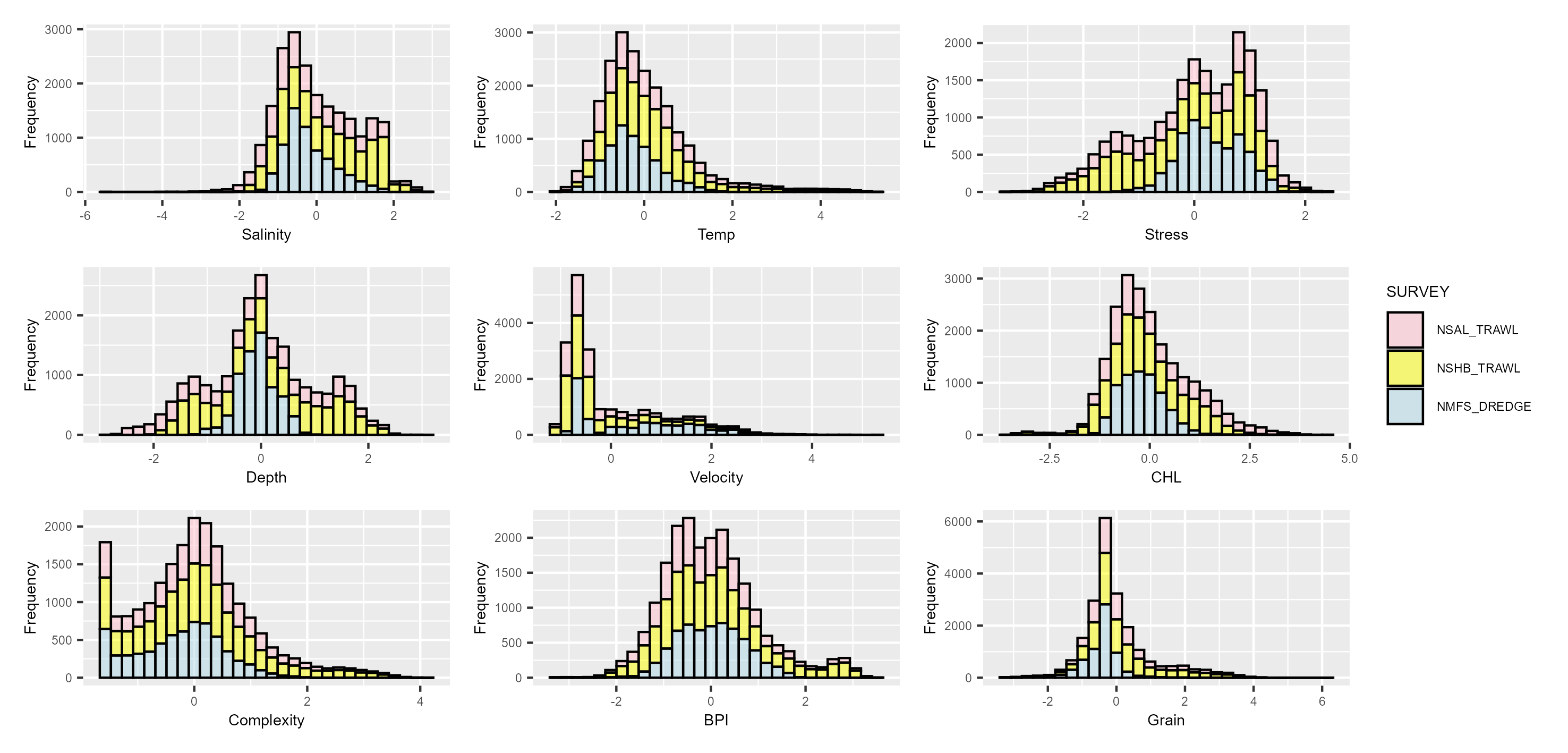}
\caption{Histograms of environmental covariates across the three surveys in the scallop abundance data. 
}
\label{fig:scallop_covariates}
\end{figure}

All covariates used for the analysis are visualized in Figure \ref{fig:scallop_covariates}.

\subsection{Additional estimation results}
\label{subsection:add.results}

\begin{figure}[ht!]
\centering
\includegraphics[width = \textwidth]{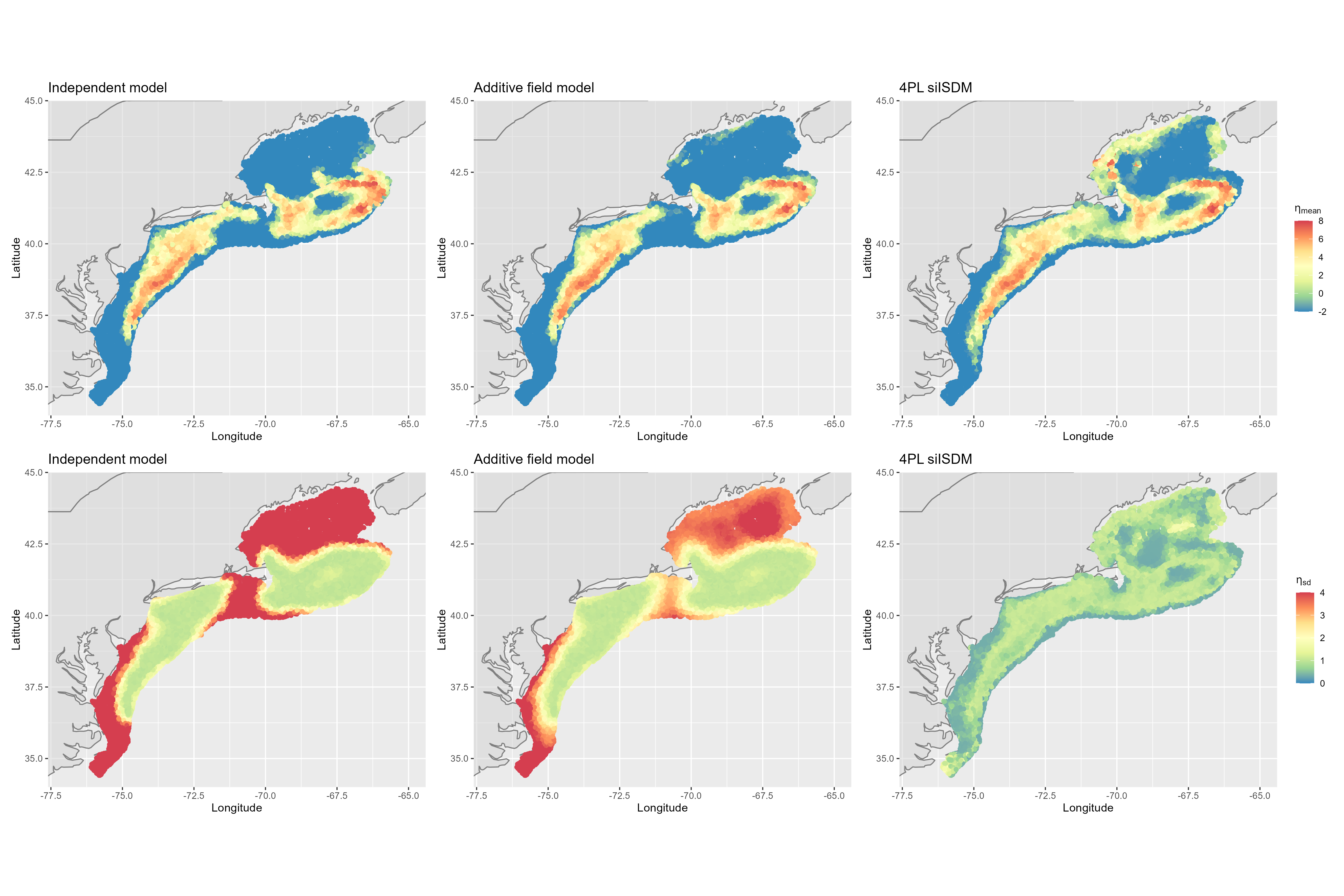}
\caption{Estimated field $\eta_3(\cdot)$ for the three models. From left to right: Independent model, Additive field ISDM and Single index ISDM. From top to bottom: Prediction mean of $\eta_3(\cdot)$ and prediction standard deviation of $\eta_3(\cdot)$.
}
\label{fig:spatial_pred}
\end{figure}

We examine the spatial distribution of species abundance. To facilitate a meaningful comparison between the models, we examine $\eta_3(\cdot)$, the expected log count for the dredge survey.
The overall spatial trend of the species distribution is captured by all models: there is a higher species distribution in the region from 75$\degree$W to 70$\degree$W, 37.5$\degree$N to 40$\degree$N, and the region from 70$\degree$W to 65$\degree$W, 40$\degree$N to 42.5$\degree$N. However, there are distinct differences in the prediction mean produced by the three models.
First, for the independent model, as there is no borrowing strength between the surveys, the estimated $\eta_2(\cdot)$ for the dredge survey cannot incorporate information from the trawl survey. For example, the region around 70$\degree$W, 42.5$\degree$N to 45$\degree$N was unobserved in the dredge survey, and is estimated to have mostly zero species count by the independent model, but they actually have non-zero counts from the trawl surveys. This also results in higher prediction standard deviation for the unobserved regions, as shown in the bottom left panel of Figure \ref{fig:spatial_pred}. 
Moving to the integrated models, we see in Figure \ref{fig:spatial_pred}, the prediction mean of $\eta_3(\cdot)$ is mostly the same for additive field ISDM and the 4PL siISDM. However, the prediction standard deviations for $\eta_3(\cdot)$ are slightly different. The reason is that the siISDM used a shared latent spatial process, instead of an added-on process, hence the prediction standard deviation is, in general, smaller for the siISDM than for the additive field ISDM, especially for regions unobserved in the dredge survey.

We also present results from fitting an additive field ISDM, in particular the predicted spatial random effects  $\tilde u_2(\cdot)$ and $\tilde u_3(\cdot)$ which represent differences from the pre-2009 bottom trawl which we used as the reference survey. We see that there are some regions where the process $\tilde u_3(\cdot)$, which represents the difference from the pre-2009 bottom trawl to the dredge, is estimated to be negative. This result is not desirable as the negative differences mean in those spatial regions, the dredge survey can have lower catch efficiency than the trawl survey.

\begin{figure}[h!]
\centering
\includegraphics[width = \textwidth]{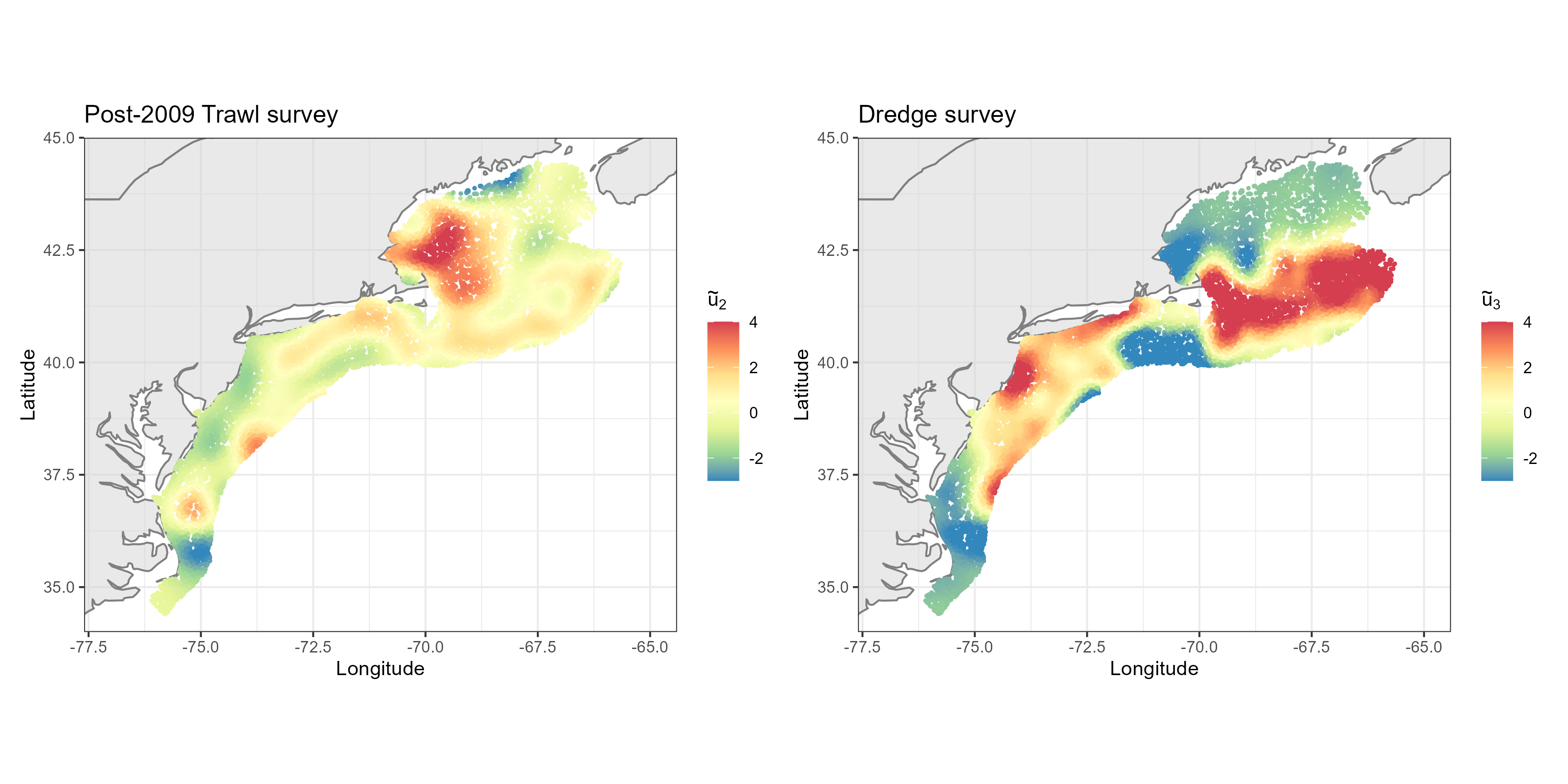}
\caption{The estimated additive fields $\tilde u_2(\cdot)$ and $\tilde u_3(\cdot)$ by the additive field ISDM. 
}
\label{fig:spatial_bias}
\end{figure}

\section{Further models details} \label{sec:isdm_details}

\subsection{Marginal log-likelihood functions} \label{sec:marginal-loglik}

We provide additional details on the marginal log-likelihood functions for the models reviewed in Section \ref{subsec:existing_models} of the main text as follows. For independent species distribution models, we have
\begin{align*}
\ell_{\text{Ind}}(\thetavec_{\text{Ind}}) = \sum_{j=1}^m \log \left\{\int \prod_{i=1}^{n_j} f\{y_{j}(\svec_i) \mid \mu_j(\svec_i), \phivec_j\} f(\alphavec_j \mid \sigma_{\alpha,j}^2, l_{\alpha,j}) f(\xivec_j \mid \sigma_{\xi,j}^2) \; \txd \alphavec_j \txd \xivec_j \right\},
\end{align*}
where $u_j(\svec_i) = \bvec(\svec_i)^\top \alphavec_j + \xi_j(\svec_i)$, the distributions of $\alphavec_j$ and $\xivec_j$ defined analogously as in $\ell_{\text{siISDM}}(\thetavec)$, and for simplicity we have used the same bisquare basis functions across all $m$ data sources although this need not be the case.
Because the log-likelihood is comprised of a sum over $m$ data sources, then independent species distribution models can be fitted in parallel across the $m$ sources. The full parameter vector is given by $\thetavec_{\text{Ind}} = (\betavec_1^\top, \ldots, \betavec_m^\top, \bm{\phi}_1^\top, \ldots, \bm{\phi}_m^\top, \sigma_{\alpha,1}^2, \ldots, \sigma_{\alpha,m}^2, l_{\alpha,1}, \ldots, l_{\alpha,m}, \sigma_{\xi,1}^2, \ldots \sigma_{\xi,m}^2)^\top$.

For the additive field ISDM, we have
\begin{align*}
\ell_{\text{AF}}(\thetavec_{\text{AF}}) &= \log \left\{\int \prod_{j=1}^m \prod_{i=1}^{n_j} f\{y_{j}(\svec_i) \mid \mu_j(\svec_i), \phivec_j\} f(\alphavec_1 \mid \sigma_{\alpha,1}^2, l_{\alpha,1}) f(\xivec_1 \mid \sigma_{\xi,1}^2) \right. \\
&\quad \left. \times \prod_{j=2}^m f(\tilde{\alphavec}_j \mid \sigma_{\alpha,j}^2, l_{\alpha,j}) f(\tilde{\xivec}_j \mid \sigma_{\xi,j}^2) \; \txd \alphavec_1 \txd \xivec_1 \prod_{j=2}^m \txd \tilde{\alphavec}_j \txd \tilde{\xivec}_j \right\},
\end{align*}
where $u_1(\svec_i) = \bvec(\svec_i)^\top \alphavec_1 + \xi_1(\svec_i)$, $\tilde{u}_j(\svec_i) = \bvec(\svec_i)^\top \tilde{\alphavec}_j + \tilde{\xi}_j(\svec_i)$ for $j = 2,\ldots,m$, and the distributions $f(\alphavec_1 \mid \sigma_{\alpha,1}^2, l_{\alpha,1}), f(\xivec_1 \mid \sigma_{\xi,1}^2), f(\tilde{\alphavec}_j \mid \sigma_{\alpha,j}^2, l_{\alpha,j})$, and $f(\tilde{\xivec}_j \mid \sigma_{\xi,j}^2)$ are defined analogously to $\ell_{\text{siISDM}}(\thetavec)$.  The full parameter vector is given by $\thetavec_{\text{AF}} = (\betavec^\top, \bm{\phi}_1^\top, \ldots, \bm{\phi}_m^\top, \sigma_{\alpha,1}^2, \ldots, \sigma_{\alpha,m}^2, l_{\alpha,1}, \ldots, l_{\alpha,m}, \sigma_{\xi,1}^2, \ldots \sigma_{\xi,m}^2)^\top$.

Finally, for the additive constant ISDM the marginal log-likelihood is given by
\begin{align*}
\ell_{\text{AC}}(\thetavec_{\text{AC}}) = \log \left\{\int \prod_{j=1}^m \prod_{i=1}^{n_j} f\{y_{j}(\svec_i) \mid \mu_j(\svec_i), \phivec_j\} f(\alphavec \mid \sigma_{\alpha}^2, l_{\alpha}) f(\xivec \mid \sigma_{\xi}^2) \; \txd \alphavec \txd \xivec \right\},
\end{align*}
which is similar is the siISDM, except the full parameter vector is given by $\thetavec_{\text{AC}} = (\beta_{10},\ldots, \beta_{m0}, \betavec_{-1}^\top, \bm{\phi}_1^\top, \ldots, \bm{\phi}_m^\top, \sigma_{\alpha}^2, l_{\alpha}, \sigma_{\xi}^2)^\top$. 

\subsection{Construction of I-splines}\label{sec:I-splines} 

In this section, we provide details on the construction of I-splines for the spline-based siISDM.
Given a set of knots $\{v_1, \dots, v_{V + k + 3} \}$, where $L = v_1 = \dots = v_{k+2} < v_{k+3} \dots < v_{V + 1} < v_{V + 2} = \dots = v_{V + k + 3} = U$, then a $k$th degree I-spline $\zetavec(x) = \{I_{k+1, 1}(x), \dots, I_{k+1, V}(x)\}^{\top}$ defined on an interval $x \in [L,U]$ is constructed recursively from a set of $(k+1)$-degree M-splines. That is,
\begin{align*}
M_{1, i}(x) &= \frac{1}{ (v_{i+1} - v_{i}) }; \; i = 1, \dots, V + 2, \\
M_{j+1, i}(x) &= \nu_{ij}  \{ (x - v_{i}) M_{j, i}(x) + (v_{i+j+1} - x) M_{j, i+1}(x) \}; \; i = 1, \dots, V + 1, j = 1, \dots, k+1, \\
I_{k+1, i-1}(x) &= \sum_{l=i}^{V+1} \frac{v_{l+k+2} - v_l}{k+2} M_{k+2, l}(x); \; i = 2, \dots, V + 1,
\end{align*}
where $\nu_{ij} = (j+1)\{j (v_{i+j+1} - v_{i} )\}^{-1}$.

\section{Additional results for numerical study}\label{sec:additional_sims}

In this section, we provide additional details and results for the numerical study.

\subsection{Additional figures}

\begin{figure}[h!]
\centering
\includegraphics[width = 0.4\textwidth]{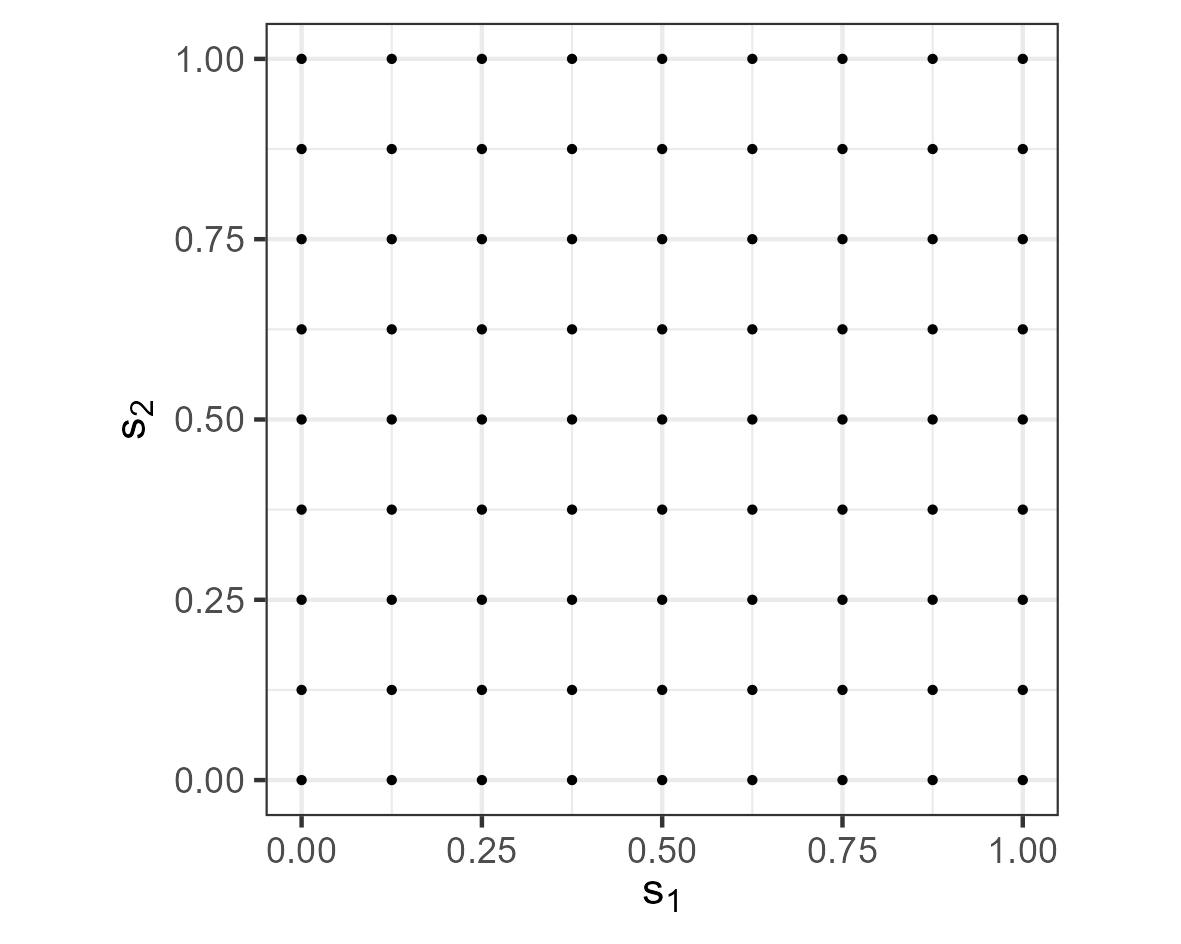}
\caption{The centroids of the basis functions used in simulating the dataset in Section \ref{sec:simulation}.
}
\label{fig:sim_centroid}
\end{figure}

\begin{figure}[h!]
\centering
\includegraphics[width = 0.7\textwidth]{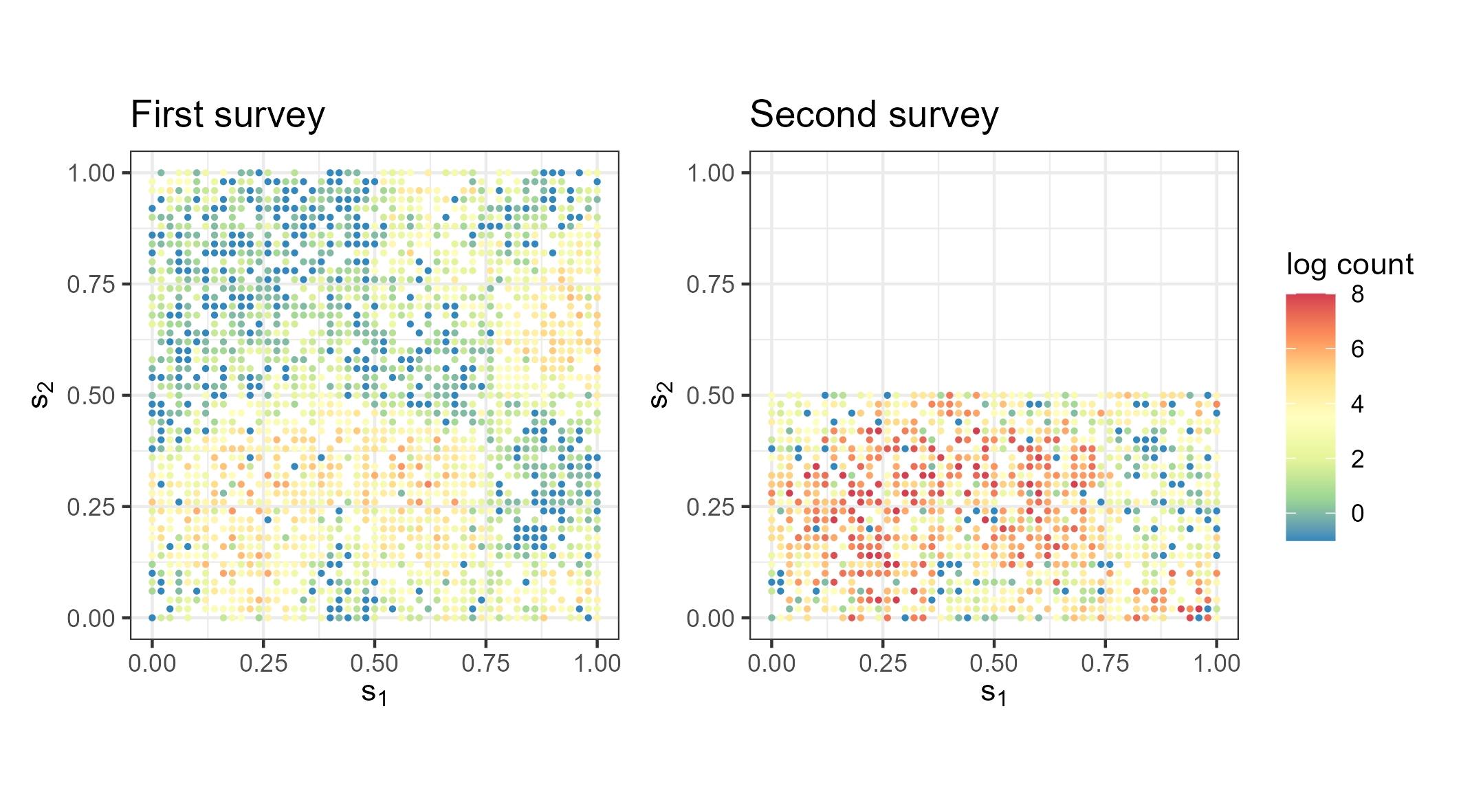}
\caption{An example of a subsampled dataset used for fitting the models in Section \ref{sec:simulation} of the main text, with species records for the first (left) and second (right) surveys. 
}
\label{fig:sim_observation_example}
\end{figure}

\begin{figure}[h!]
\centering
\includegraphics[width = 0.9\textwidth]{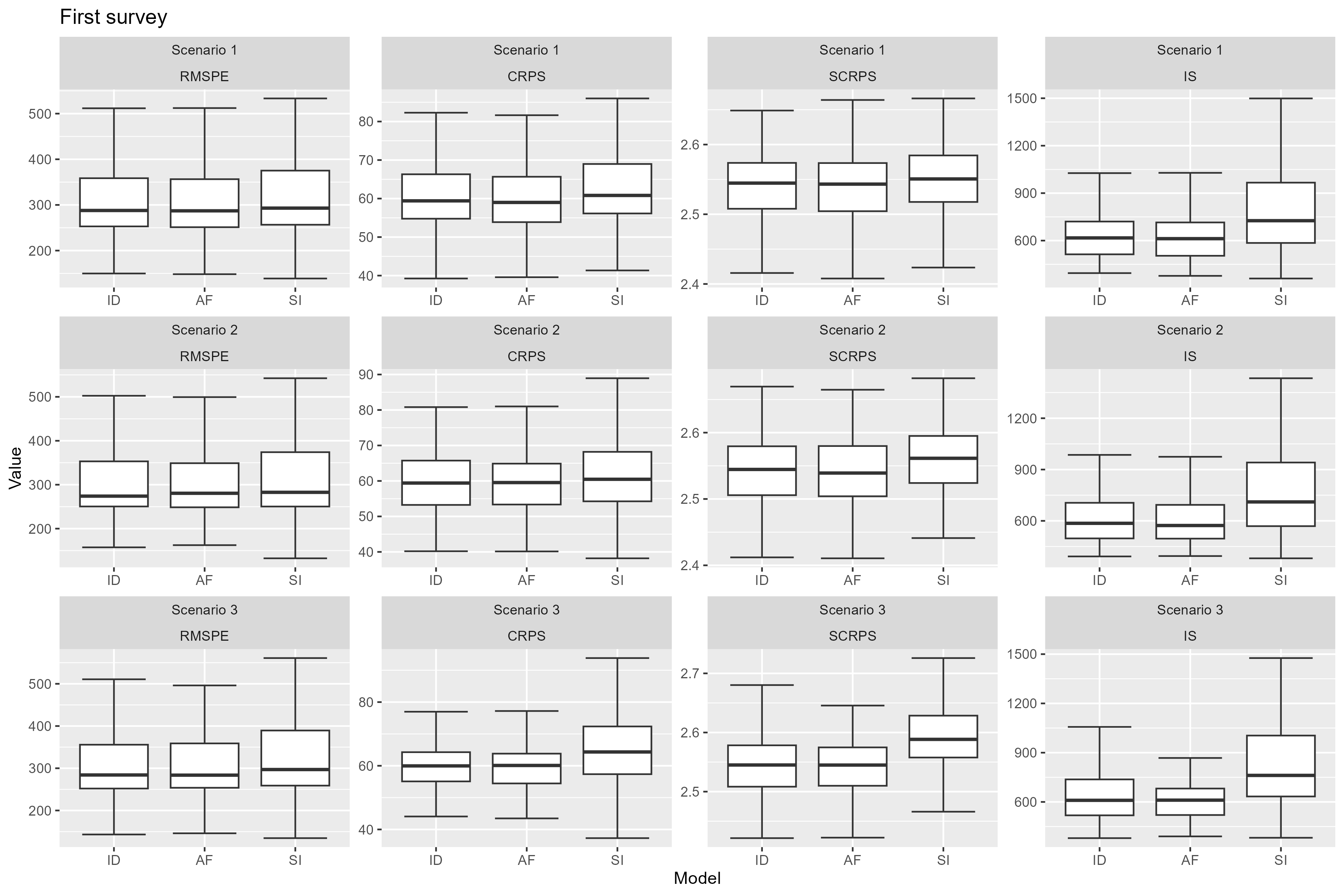}
\includegraphics[width = 0.9\textwidth]{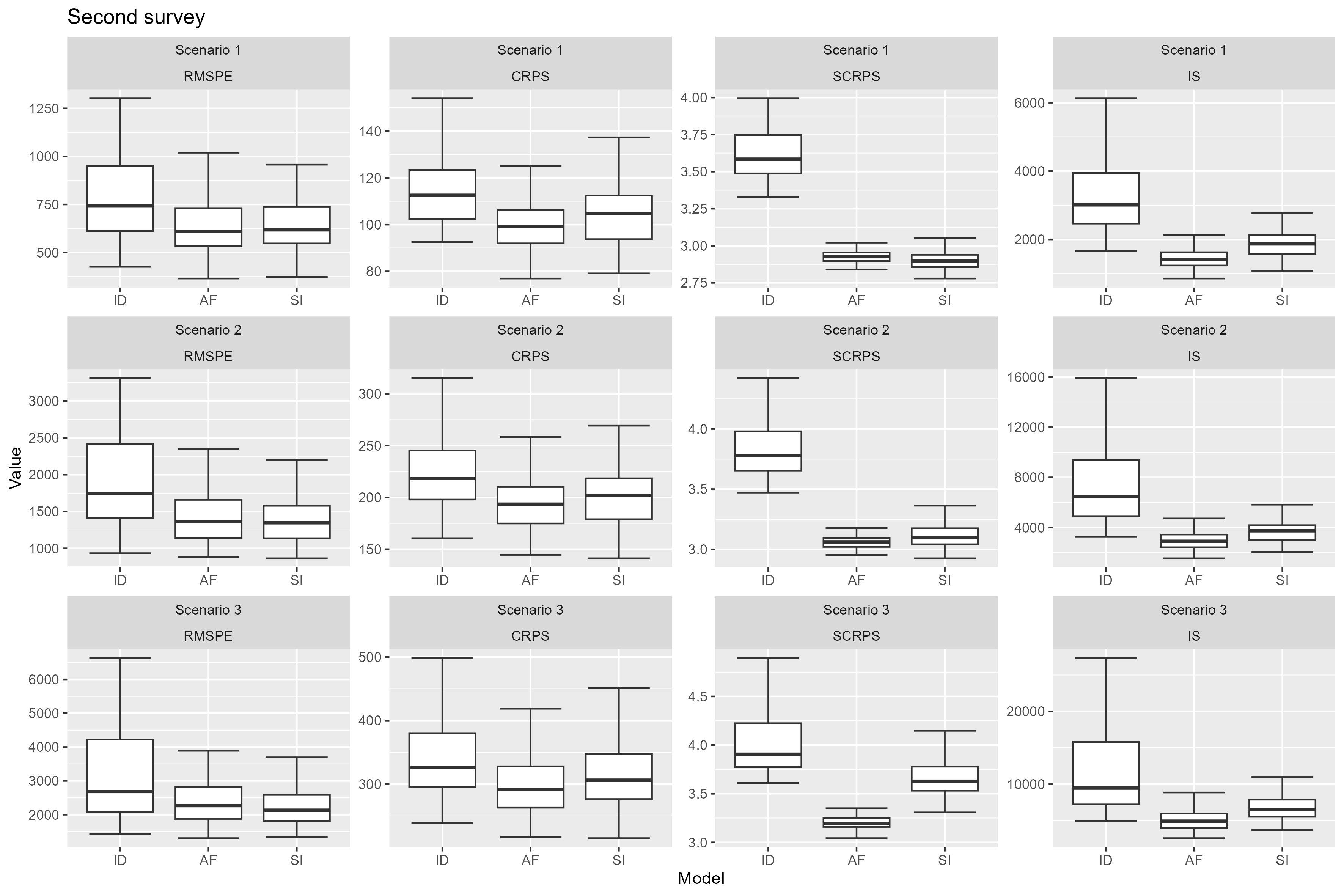}
\caption{Boxplots of the prediction metrics for datasets simulated from the siISDMs.
The rows represent Case 1, Case 2 and Case 3 for the first survey and the second survey.
The columns represent the RMSPE, CRPS, SCRPS and IS metrics across 100 simulated datasets for each case.
}
\label{fig:sim1_prediction_metrics}
\end{figure}

\FloatBarrier

\subsection{Additional details for prediction metrics}

Consider the simulation design in Section \ref{sec:simulation}, and suppose at spatial location $\svec_i$ in a test dataset we have survey-specific recorded abundance $\tilde y_j(\svec_i); j = 1,2$. Let $\hat y_j(\svec_i)$ generically denote the predicted abundance from one of the three methods considered i.e., siISDM, independent species distribution models, and additive field ISDM. Then we consider the following metrics:

\begin{itemize}
    \item Root mean squared error of prediction, which is given by  $\text{RMSEP}_j = \left[n^{-1}_{j,\text{test}} \sum_{i = 1}^{n_{j,\text{test}}} \{\hat y_j(\svec_i) - \tilde y_j(\svec_i)\}^2\right]^{1/2}$, where $n_{j,\text{test}}$ denotes the number of observations/spatial locations in the test set for survey $j$;

    \item Continuous rank probability score \citep[CRPS,][]{gneiting2007strictly}, which at a given spatial location for a count response is computed as 
    $
    \text{CRPS}_j(\svec_i) = \int \left[\hat{F}\{y_j(\svec_i)\} - I \{ y_j(\svec_i) \geq \tilde y_j(\svec_i) \} \right]^2 
    \text{d} y_j(\svec_i)
    $, 
    where $I(\cdot)$ is the indicator function and $\hat{F}(\cdot)$ is the empirical cumulative distribution function obtained from the sampling approach discussed at the end of Section \ref{subsec:estimationprediction}. The final metric is obtained by computed the CRPS at each of the $n_{j,\text{test}}$ spatial location and then averaging over the total number of spatial locations;

    \item Scaled CRPS \citep{bolin2023local} also takes into account the scale of the prediction;

    \item Interval score \citep{gneiting2007strictly} which at a given spatial location is computed as $
    \text{IS}_j(\svec_i) = \{u_j(\svec_i) - l_j(\svec_i)\} + \frac{2}{\alpha} \{l_j(\svec_i) - \tilde y_j(\svec_i)\} I\{\tilde y_j(\svec_i) < l_j(\svec_i)\} + \frac{2}{\alpha} \{\tilde y_j(\svec_i) - u_j(\svec_i)\} I\{\tilde y_j(\svec_i) > u_j(\svec_i)\}
$,
where $\{l_j(\svec_i), u_j(\svec_i)\}$ generally denote the $ 100 (1-\alpha)\% $ prediction intervals for $y_j(\svec_i)$ e.g., in the siISDM it is obtained as a percentile interval based on the sampling approach discussed at the end of Section \ref{subsec:estimationprediction}, while an analogous approach can be adopted for the independent species distribution model and the additive field ISDM.
\end{itemize}

\subsection{Simulation from additive field ISDM}
\label{subsection:simulation-additive}

In this section, we provide an additional numerical study with data generated from the additive field ISDM.

We generate simulated data from the additive field ISDM as follows. The covariates and spatial random effect for the reference survey (i.e., $u_1(\cdot)$) are simulated in the same way as in Section \ref{sec:simulation}.
Set the intercepts for the two surveys to be $(3, 4)^{\top}$, 
Next, following the formulation in Section \ref{subsec:existing_models}, we set 
$u_2(\cdot) = u_1(\cdot) + \tilde u_2(\cdot)$ where $\tilde u_2(\cdot) = \bvec(\cdot) \tilde \alphavec_2$ and $\tilde \alphavec_2$ is simulated from a multivariate normal distribution with an exponential covariance function, using length equal to $l_{\alpha,2} = 0.05$. Again, note the true data generation process here does not include a fine-scale term $\xi(\cdot)$. On the other hand, we consider three cases for the variance of $\alphavec_2$, with $\sigma^2_{\alpha,2} = \{0.1, 0.2, 0.5\}$; a smaller variance can be interpreted as the second survey having a mean recorded species abundance that is more similar to that of the first survey (up to a scaling factor), while a larger variance means the recorded species abundances from the two surveys are more different e.g., their characteristic gears/sampling protocols are more distinct. Finally, the remainder of the set and the data are sampled in the same way as in Section \ref{sec:simulation} e.g., we simulate count responses from a negative binomial distribution with survey-specific dispersion parameters $\phi_1 = 1$ and $\phi_2 = 0.5$.

\begin{figure}[h!]
\centering
\includegraphics[width = 0.9\textwidth]{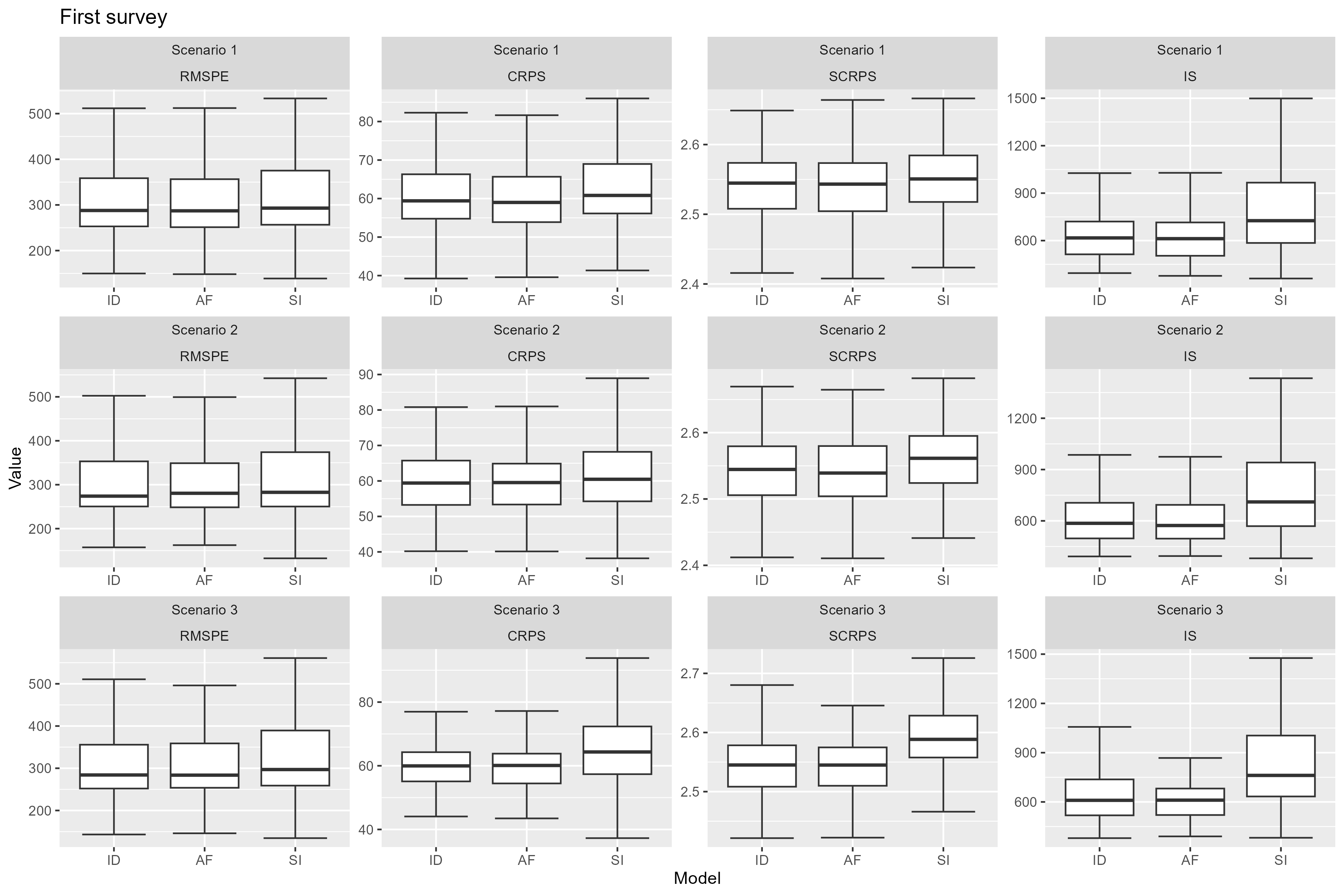}
\includegraphics[width = 0.9\textwidth]{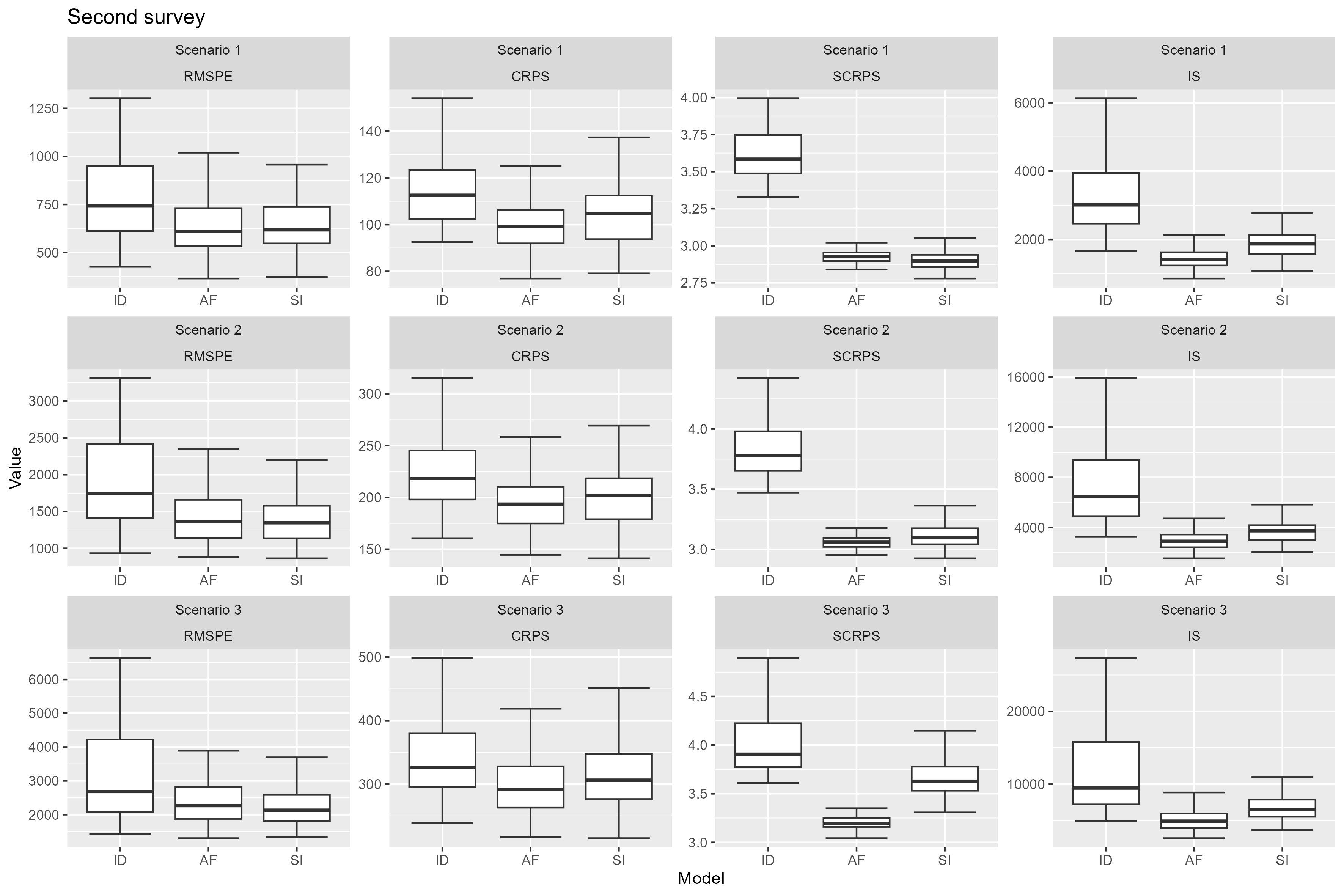}
\caption{Boxplots of the prediction metrics for datasets simulated from the additive field ISDMs. 
The rows represent Case 1, Case 2 and Case 3 for the first survey and the second survey.
The columns represent the RMSPE, CRPS, SCRPS and IS metrics across 100 simulated datasets for each case.
}
\label{fig:sim2_prediction_metrics}
\end{figure}

Figure \ref{fig:sim2_prediction_metrics} presents comparative boxplots for the predictive metrics for the two surveys obtained from fitting the three models to 100 simulated datasets, and for each of the three cases of $\sigma^2_{\alpha, 2}$. For cases 1 and  2 where the variance of the additive field in the second survey is smaller, the siISDM can produce similar to or slightly worse metrics than the additive field ISDM, even though it is a severely misspecified model. For case 3 where the variance of the additive field is larger, the predictive performance of the siISDM deteriorates especially for the first survey, while for the second survey it still performs better than independent species distribution models.


\end{document}